\documentclass[pre,preprint,superscriptaddress, aps]{revtex4-2}

\usepackage{amsmath}
\usepackage{amssymb}
\usepackage{graphicx}
\usepackage{verbatim}
\usepackage{xcolor}

\newcommand{\be}{\begin{equation}}
\newcommand{\ee}{\end{equation}}

\newcommand {\lb}{\langle}
\newcommand {\rb}{\rangle}
\newcommand{\mf}{mean-field}
\newcommand{\mft}{mean-field theory}
\newcommand{\lam}{\lambda}
\newcommand{\e}{\epsilon}

\newcommand{\wmax}{w_{\max}}
\newcommand{\ebar}{\lb E \rb}
\newcommand{\trich}{\tau_{\rm ra}}

\begin{document}

\title{Simulation of a generalized asset exchange model with economic growth and wealth distribution}

\author{Kang K. L. Liu}

\affiliation{Department of Physics, Boston University, Boston, Massachusetts 02215}

\author{N. Lubbers}
\altaffiliation[Present address: ]{Los Alamos National Laboratory, Los Alamos, New Mexico 87545}
\affiliation{Department of Physics, Boston University, Boston, Massachusetts 02215}

\author{W. Klein}

\affiliation{Department of Physics, Boston University, Boston, Massachusetts 02215}
\affiliation{Center for Computational Science, Boston University, Boston, Massachusetts 02215}

\author{J. Tobochnik}
\affiliation{Department of Physics, Kalamazoo College, Kalamazoo, Michigan 49006}
\affiliation{Department of Physics, Boston University, Boston, Massachusetts 02215}

\author{B. M. Boghosian}
\affiliation{Department of Mathematics, Tufts University, Medford, Massachusetts 02155}

\author{Harvey Gould}
\email{hgould@clarku.edu}
\affiliation{Department of Physics, Boston University, Boston, Massachusetts 02215}
\affiliation{Department of Physics, Clark University, Worcester, Massachusetts 01610}

\date{\today}

\begin{abstract}
The agent-based Yard-Sale model of wealth inequality is generalized to incorporate exponential economic growth and its distribution. The distribution of economic growth is nonuniform and is determined by the wealth of each agent and a parameter $\lam$. Our numerical results indicate that the model has a critical point at $\lam=1$ between a phase for $\lam < 1$ with economic mobility and exponentially growing wealth of all agents and a non-stationary phase for $\lam \geq 1$ with wealth condensation and no mobility. We define the energy of the system and show that the system can be considered to be in thermodynamic equilibrium for $\lam < 1$. 
Our estimates of various critical exponents are consistent with a mean-field theory (see following paper). The exponents do not obey the usual scaling laws unless a combination of parameters that we refer to as
the Ginzburg parameter is held fixed as the phase transition is approached. The model illustrates that both poorer and richer agents benefit from economic growth if its distribution does not favor the richer agents too strongly. This work and the following theory paper contribute to our understanding of whether the methods of equilibrium statistical mechanics can be applied to economic systems.
\end{abstract}


\maketitle

\section{Introduction and the GED model}

Although economies are complex systems that are difficult to understand~\cite{review}, the consideration of simple models can provide insight if the questions are of a statistical nature and about the economy as a whole. 
One such question is whether economic growth can benefit all members of society~\cite{review, gud, jfk}. Another question is to what degree is wealth accumulation a natural consequence of the way that wealth is exchanged and distributed. Whether these questions and others can be treated using methods appropriate to equilibrium systems is not settled~\cite{ole}.

In this paper we approach these questions by simulating an agent-based model that incorporates wealth exchange, economic growth, and its distribution. Agent-based wealth-exchange models
exhibit a rich phenomenology~\cite{angle, angle2, saving, money, Moukarzel, rmp, ajp, Bouchaud}. The agent-based asset-exchange model we generalize belongs to a class of wealth exchange models that has been of considerable interest and has provided insight into how economies function~\cite{Burda, boghos, boghos2, boghos3, Chorro, krapiv, ispo, wealthattained, devitt, boghos4, sciamer}. In these models the amount of wealth transfer is determined stochastically to represent the uncertainty of the value of an agent's assets.

A particularly interesting agent-based model that incorporates wealth exchange is the Yard-Sale model~\cite{money, rmp, saving, boghos, ijmp, hayes, krapiv, sciamer, Burda, devitt, Moukarzel, Chorro, boghos4, wealthattained, ajp, Bouchaud, boghos2, boghos3, angle, angle2, ispo}. In this model two agents are chosen at random and a fraction $f$ of the wealth of the poorer agent is transferred to the winning agent. The latter is determined at random with probability 1/2. If an equal amount of wealth is initially assigned to the $N$ agents, the result is that after many exchanges, the wealth is concentrated among increasingly fewer agents, culminating, in the limit of infinitely many wealth exchanges and $N \to \infty$, to a single agent holding a finite fraction of the total wealth~\cite{boghos, angle, ispo, krapiv, Moukarzel, boghos2, Chorro, sciamer, hayes}.

To investigate the effect of economic growth on the wealth distribution, we generalize the Yard-Sale model so that the wealth $\mu W(t)$ is added to the system after $N$ exchanges, where $W(t)$ is the total wealth at time $t$ and the time $t$ is defined such that one unit of time corresponds to $N$ exchanges; $W(t)$ grows exponentially with the rate $\mu>0$. The motivation for this type of growth is the expected annual increase in the gross domestic product~\cite{Slanina, wealthcomment}.

To distribute the increase of the total wealth due to growth to individual agents, we introduce the distribution parameter $\lam \geq 0$ and assign the added wealth to agent $i$ according to their wealth $w_i(t)$ at time $t$ as
\be
\label{distributionofgrowth}
\Delta w_{i}(t) = \mu W(t) \dfrac{w^{\lam}_{i}(t)}{\sum_{i=1}^{N} w^{\lam}_{i}(t)}.
\ee
The form of Eq.~\eqref{distributionofgrowth} implies that as $\lambda$ increases, the allocation of the added wealth is weighted more toward agents with greater wealth. 
We will refer to the model with the incorporation of exponential growth of the total wealth and the $\lam$-dependent distribution mechanism in Eq.~\eqref{distributionofgrowth} as the {\it Growth, Exchange, and Distribution} (GED) model.

The motivation for the form of the wealth distribution in Eq.~\eqref{distributionofgrowth} is that in practice, not all agents benefit equally from economic growth, and that agents with more assets and resources are able to take more advantage of the growth of the economy. We argue in the Appendix that the allocation of growth according to Eq.~\eqref{distributionofgrowth} is consistent with economic data.

The distribution parameter $\lam$, exchange parameter $f$, growth parameter $\mu$, and the number of agents $N$ determine the wealth distribution in the model. Our primary results can be grouped into two categories -- the 
implications for economic systems and the implications for our understanding of the statistical mechanics of 
systems that are near-mean-field. We find that there is a phase transition at $\lam=1$ such that for $\lam < 1$, all agents benefit from economic growth, there is economic mobility, and the system is in thermodynamic equilibrium. In contrast, for $\lam \geq 1$, the system is non-stationary, there is no economic mobility, and there is wealth condensation as in the original Yard-Sale model. In the context of statistical mechanics we note that we can define an energy that satisfies the Boltzmann distribution for $\lam < 1$. The existence of the latter is consistent with the assumption that that the system is not just in a steady state, but is in thermodynamic equilibrium for $\lam < 1$.

The remainder of the paper is structured as follows: In Sec.~\ref{sec:steady} we show that the wealth distribution reaches a steady state and that wealth condensation is avoided for $\lam < 1$. In Sec.~\ref{sec:dynamics} we 
show that the GED model is effectively ergodic and that there is economic mobility for $\lam < 1$. In Sec.~\ref{sec:equilibrium} we find that we can define 
an energy that satisfies the Boltzmann distribution. We characterize the phase transition at $\lambda = 1$  for a fixed number of agents in Sec.~\ref{sec:fixedagents} and estimate the critical exponents $\beta$, $\gamma$, and $\alpha$ associated with the behavior of the order parameter, its variance, and the energy, respectively. The result is that the exponents determined for fixed $N$ do not satisfy the usual scaling laws. The consequences of the system being describable by a mean-field theory~\cite{kleinmf} are discussed in Sec.~\ref{sec:mft}, where we introduce the Ginzburg parameter and show that scaling is restored if the transition is approached at fixed Ginzburg parameter rather than for fixed $N$. In Sec.~\ref{sec:times} we show that there is critical slowing down consistent with the mean-field theory predictions of Ref.~\cite{kleinmf}. We summarize and discuss our results in Sec.~\ref{sec:discussion}. In the appendix we argue from economic data that our method of distributing the growth is a reasonable zeroth order approximation.

\section{\label{sec:steady}Steady state wealth distribution}

\begin{figure}[tbp] 
\includegraphics[scale=0.65]{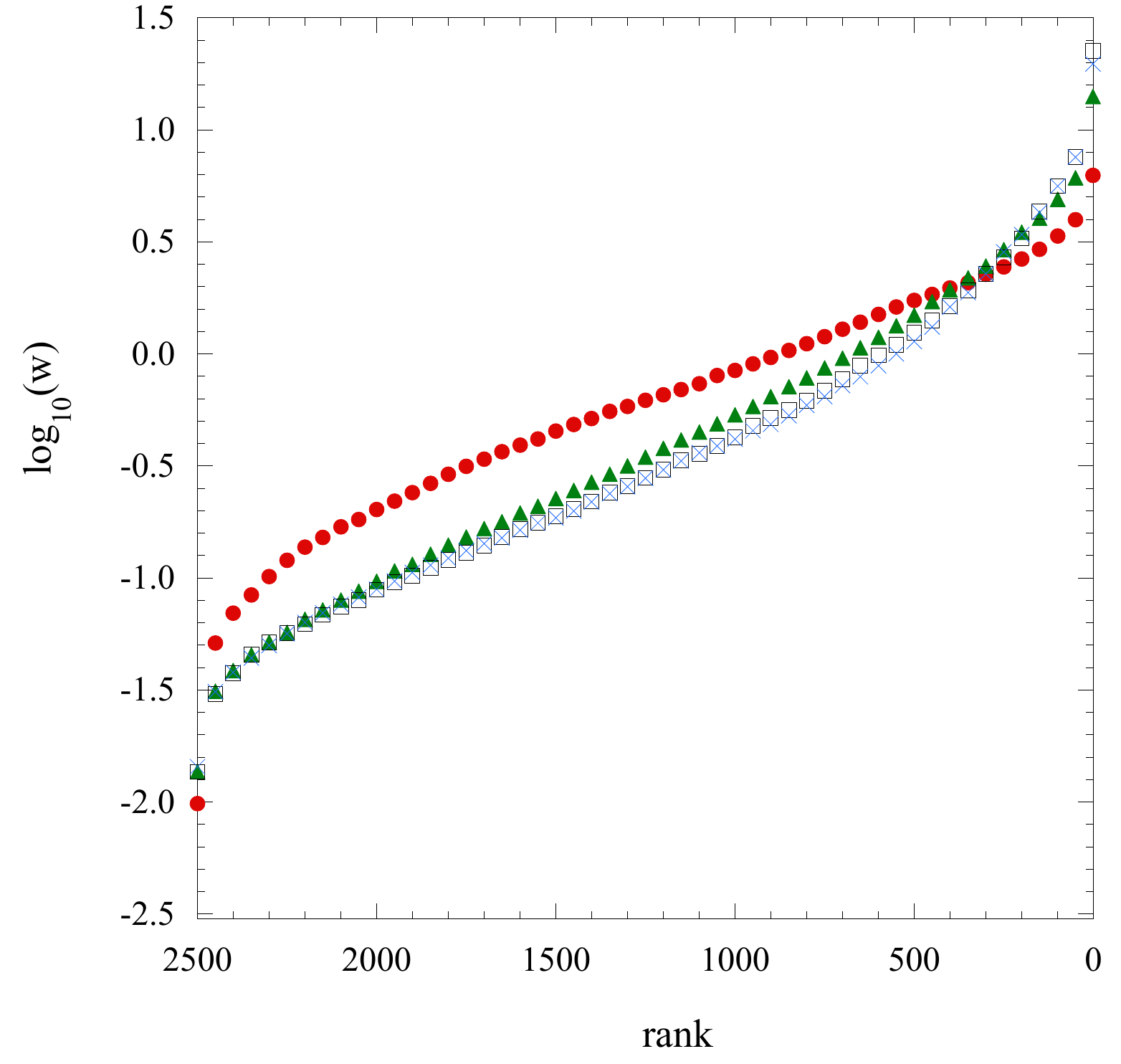}
\vspace{-0.3cm}
\caption{\label{fig:lam0}The time dependence of the rescaled wealth distribution as a function of the rank of $N=2500$ agents for $\lam=0$ at $t=100$ ($\color{red}\bullet$), $t=400$ ($\color{green}\blacktriangle$), $t=1600$ ($\square$), and $t=6400$ ($\color{blue} \times$). For $t>1600$, the rescaled wealth distributions collapse onto the same curve indicating that the rescaled wealth distribution has reached a steady state. Each agent is initially assigned wealth one, so that the total wealth is $N$ ($N=2500$, $f = 0.1$, $\mu=0.001$).} 
\end{figure}

Because the total wealth increases exponentially, it is convenient to introduce the rescaled wealth of an agent,
\be
\widetilde w_i(t) = \frac{N}{W(t)}w_i(t), \label{eq:rescale}
\ee
and consider the rescaled wealth distribution of the $N$ agents. That is, after the increased wealth due to growth is distributed to the agents, their wealth is scaled so that the total rescaled wealth equals $N$, the initial total wealth. 
In the following all references to the wealth of the agents will be to their rescaled wealth, and we will omit the tilde for simplicity.

The simulation of the GED model proceeds as follows:

\begin{enumerate}
\setcounter{enumi}{-1}
\item Usually, we assign the wealth of each agent at random and then rescale $w_i$ so that the total wealth $\sum_i w_i(t=0)$ is equal to $N$. The results for this section only are for $w_i(t=0) = 1$.

\item Choose agents $i$ and $j$ at random regardless of their wealth and determine the amount $f \min[w_i(t), w_j(t)]$ to be exchanged. Choose at random which agent gains and which agent loses. \label{step}

\item Implement step one $N$ times. Because the $N$ exchanges conserve the total wealth, the latter remains equal to $N$.

\item Assign the additional wealth  due to growth to the agents according to Eq.~\eqref{distributionofgrowth}.

\item Rescale $w_i(t)$ so that $W(t) = \sum w_i(t) = N$.

\item Set $t = t+1$.

\item Repeat steps 2--5 until a steady state wealth distribution is attained (for $\lam < 1$) and then determine the average values of the desired quantities of interest.

\end{enumerate}

The simulations in this section are for $N=2500$, $f = 0.1$, $\mu=0.001$, and various values of the distribution parameter $\lam$. The qualitative results discussed here do not depend on the values of the number of agents $N$, the fraction of the poorer agents's wealth that is exchanged $f$, and the growth parameter $\mu$. 

We show in Fig.~\ref{fig:lam0} the time dependence of the rescaled wealth distribution for $\lam=0$, starting from the initial condition $w_i(t=0) = 1$ for all $i$. The wealth disparity between richer and poorer agents initially increases until a steady state is established. Once a steady state is reached, the rescaled wealth distribution remains fixed, and the wealth in every rank increases as $e^{\mu t}$.

\begin{figure}[tbp] 
\centering
\includegraphics[scale=0.65]{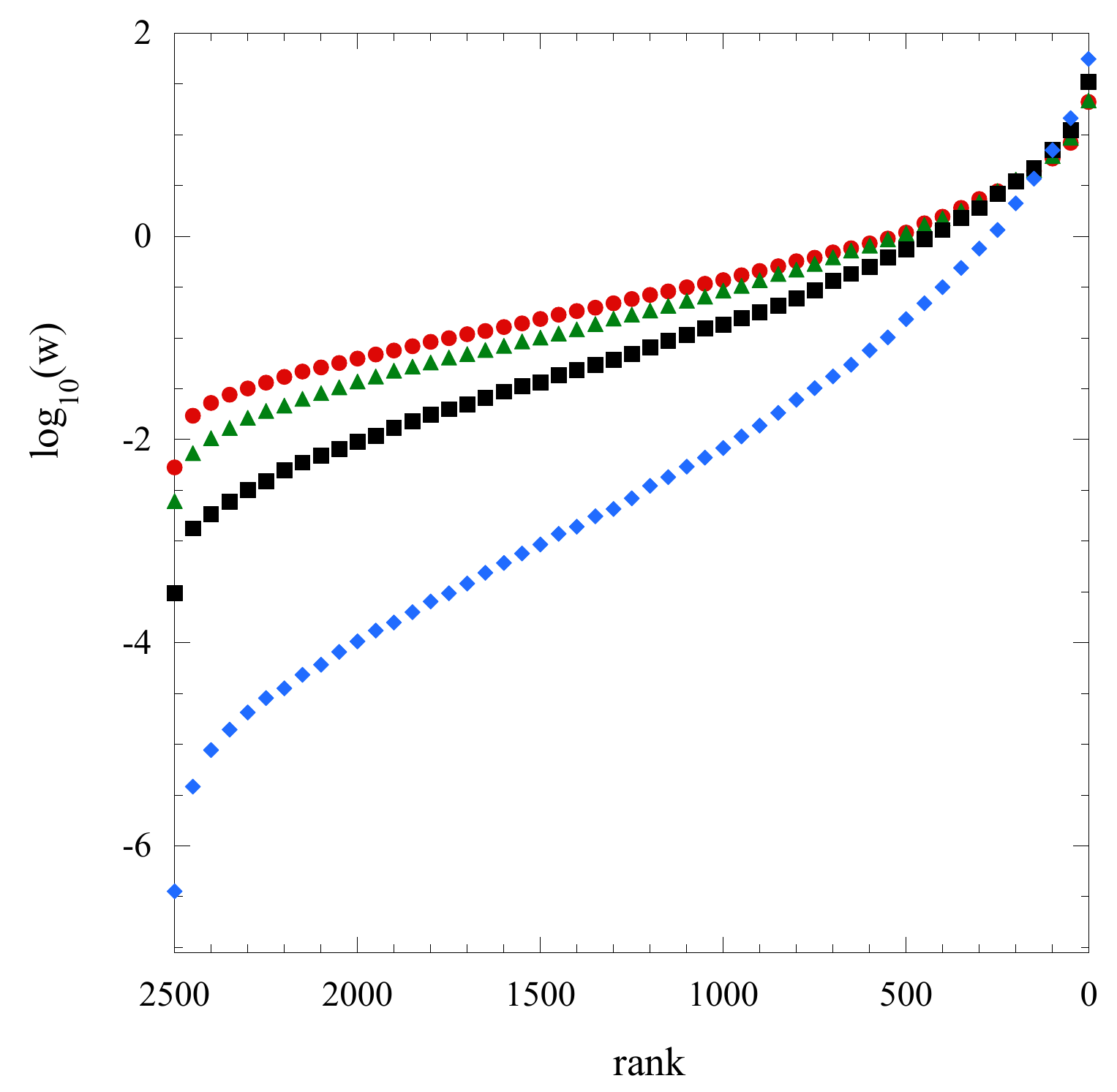}
\vspace{-0.2in}
\caption{\label{fig:differentlam}
The rescaled wealth distribution versus the rank of $N=2500$ agents for $\lam = 0.2$ ($\color{red}\bullet$), $\lam=0.4$ ($\color{green}\blacktriangle$), $\lam=0.6$ ($\blacksquare$), and $\lam=0.8$ ($\color{blue}\blacklozenge$) at $t=10^6$ after a steady state has been reached. The wealth distribution becomes less equal as $\lam \to 1^-$ ($f = 0.1$, $\mu=0.001$).}
\end{figure}

According to Eq.~\eqref{distributionofgrowth}, the growth allocation is weighted more toward the richer agents as $\lam\to 1^-$, thus leading to a less equal steady state rescaled wealth distribution (see Fig.~\ref{fig:differentlam}). The time to reach a steady state increases as $\lam$ approaches $1$, and as for $\lam=0$, the wealth of all agents increases exponentially after a steady state has been reached. (The limit $\lam \to 1$ will always be from below unless otherwise specified.) We find that for $0 \leq \lam < 1$, ``a rising tide lifts all boats'' and all agents benefit from economic growth.

\begin{figure}[t] 
\centering
\includegraphics[scale=0.65]{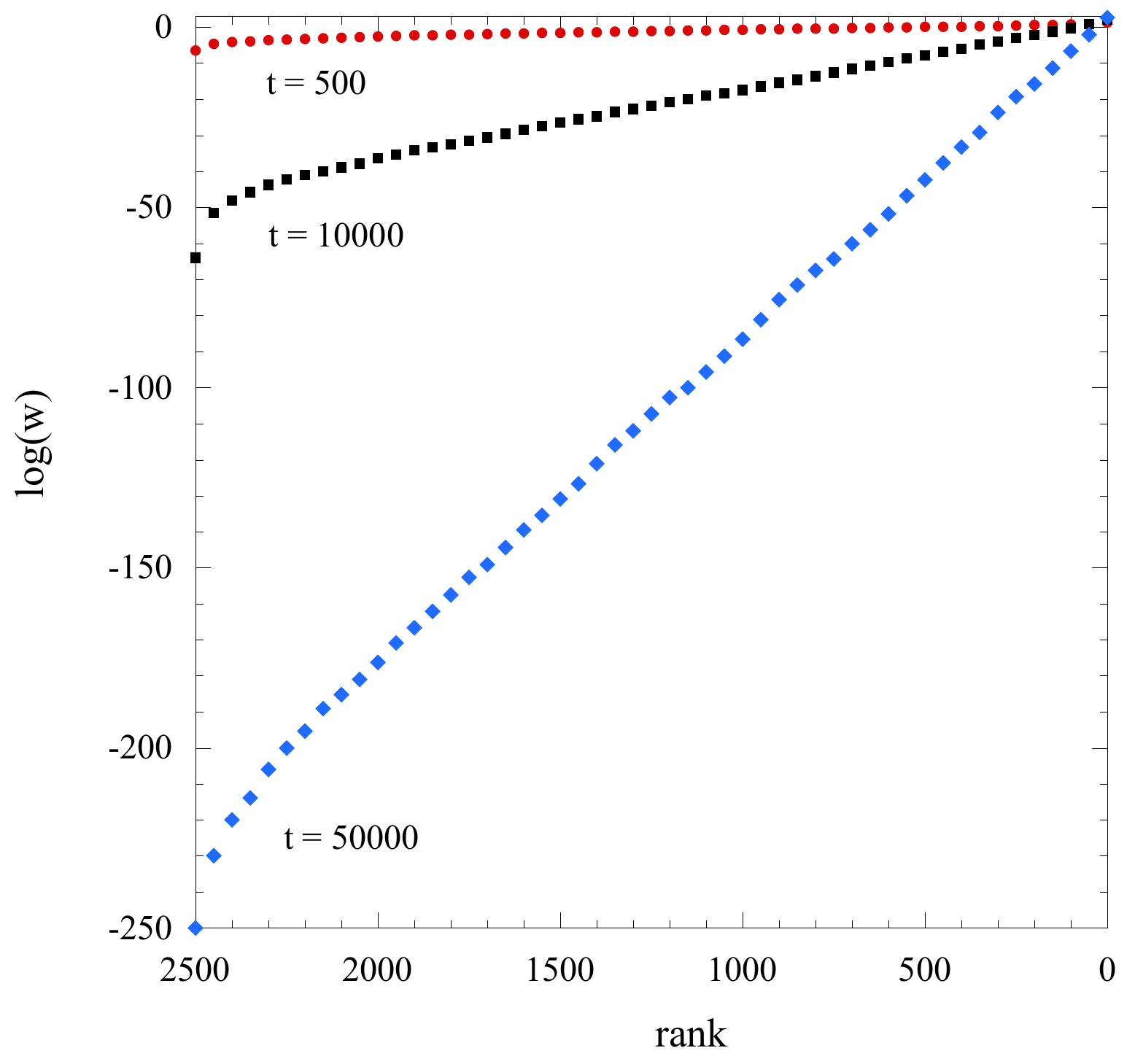}
\vspace{-0.2cm}
\caption{\label{fig:unstable}The rescaled wealth distribution at $t=500$ ($\color{red}\bullet$), $t=10000$ ($\blacksquare$), and $t=50000$ ($\color{blue}\blacklozenge$) for $\lam=1$. In contrast to the behavior for $\lam < 1$, the slope of the rescaled wealth distribution decreases with time and a steady state is not reached. The time for a single agent to gain almost all of the wealth is a decreasing function of $\lam$ for $\lam>1$ ($N=2500$, $f = 0.1$, $\mu=0.001$).}
\end{figure}

The time-dependence of the rescaled wealth distribution is shown for $\lam = 1$ in Fig.~\ref{fig:unstable}. A steady state is not reached, and the slope of the rescaled wealth distribution increases with time, corresponding to the accumulation of wealth by fewer and fewer agents until eventually a single agent gains almost all the wealth. Similar results are found for $\lam > 1$. The time for a single agent to dominate decreases as $\lam$ increases for $\lam>1$. 

Note that $\lam=1$ is a special case for which the increase in an agent's wealth due to growth is proportional to the agent's wealth. Hence, for $\lam=1$, the ratio of the rescaled wealth of any two agents does not change after the distribution of the growth in wealth according to Eq.~\eqref{distributionofgrowth}. Consequently, aside from the exponential growth of the total wealth, the entire dynamics of the system is driven only by the wealth exchange mechanism, and the evolution of the wealth distribution for $\lam = 1$ is identical to its evolution in the original Yard-Sale model; that is, the model with no economic growth.

\section{\label{sec:dynamics}Effective ergodicity and economic mobility}

The numerical results in Sec.~\ref{sec:steady} indicate that the GED model exhibits distinct behavior for $\lam < 1$ and $\lam \geq 1$. In particular, for $\lam<1$, all agents benefit from economic growth, whereas for $\lam \geq 1$ only the richest agent becomes richer. In Sec.~\ref{sec:ergodic} we show that the GED model is effectively ergodic for $\lam < 1$, but is not ergodic for $\lam \geq 1$. In Sec.~\ref{sec:mobility} we find that the agents have nonzero economic mobility for $\lam < 1$, but have zero mobility for $\lam \geq 1$. Unlike in Sec.~\ref{sec:steady}, we randomly assign the wealth of each agent at $t=0$ from a uniform distribution and then rescale the agents' wealth so that the initial total wealth equals $N$.

\subsection{\label{sec:ergodic}Effective ergodicity}
To determine whether the system is effectively ergodic, we define
the (rescaled) wealth metric as~\cite{tm}
\be
\label{eq:metric}
\Omega(t) = \frac{1}{N}\sum_{i=1}^{N}\big [ {\overline w}_{i}(t) - \overline{w}(t) \big]^{2},
\ee
where $\overline{w}_{i}(t)$ is the time averaged wealth of agent $i$ at time $t$,
\begin{align}
\overline{w}_{i}(t) &= \frac{1}{t}\!\int_{0}^{t} w_{i}(t')\,dt', \\
\noalign{\noindent and $\overline{w}(t)$ is the average over all agents,}
\overline{w}(t) &= \frac{1}{N} \sum_{i=1}^{N}{\overline w}_{i}(t).
\end{align}
The metric $\Omega(t)$ in Eq.~\eqref{eq:metric} is a measure of how the time averaged  wealth of each agent approaches the  wealth averaged over all agents.
If the system is effectively ergodic, $\Omega(t)\propto 1/t$~\cite{tm}. Effective ergodicity is a necessary, but not a sufficient condition for ergodicity.

\begin{figure}[tbp]
\centering
\includegraphics[scale=0.48]{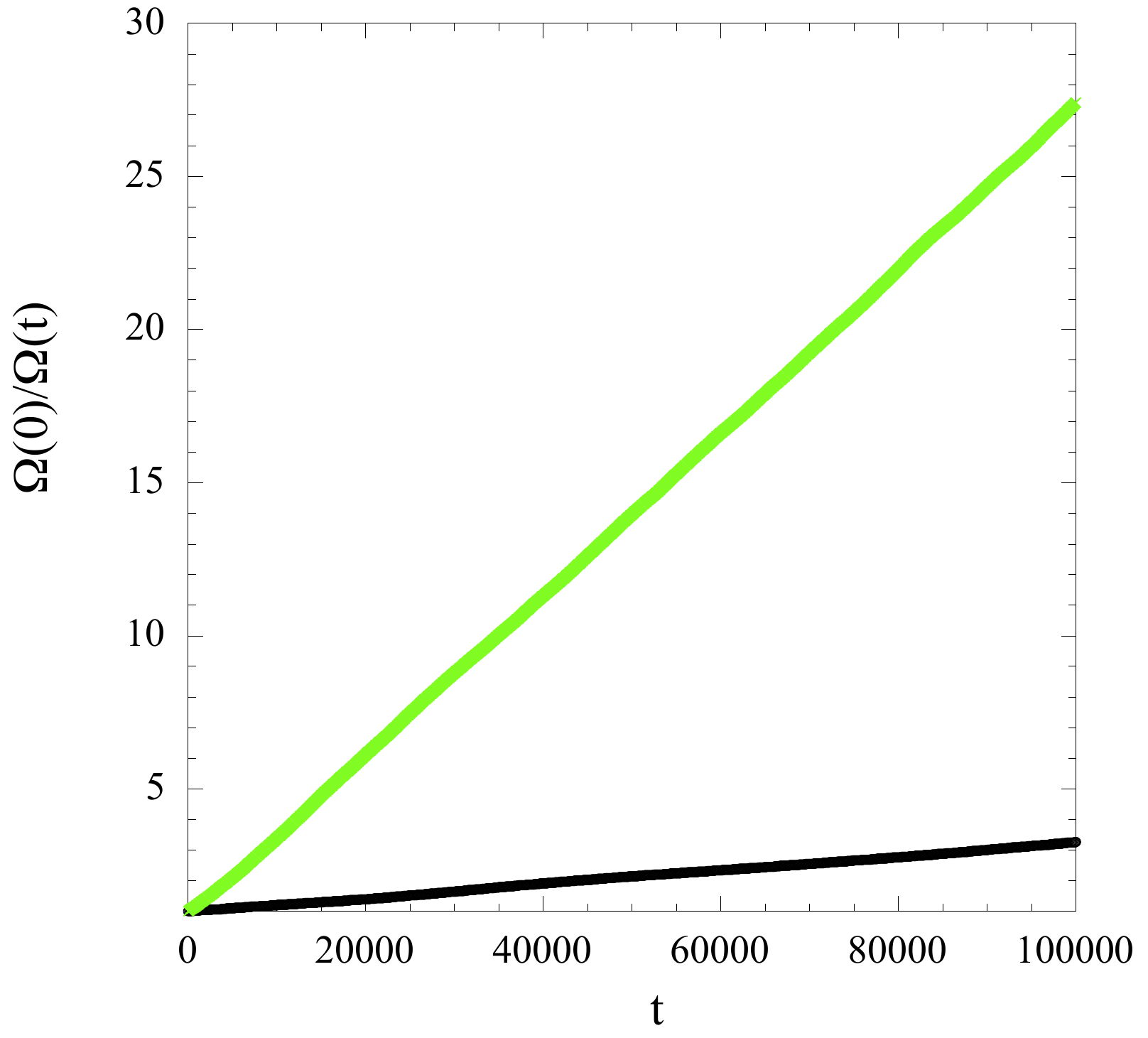}
\includegraphics[scale=0.48]{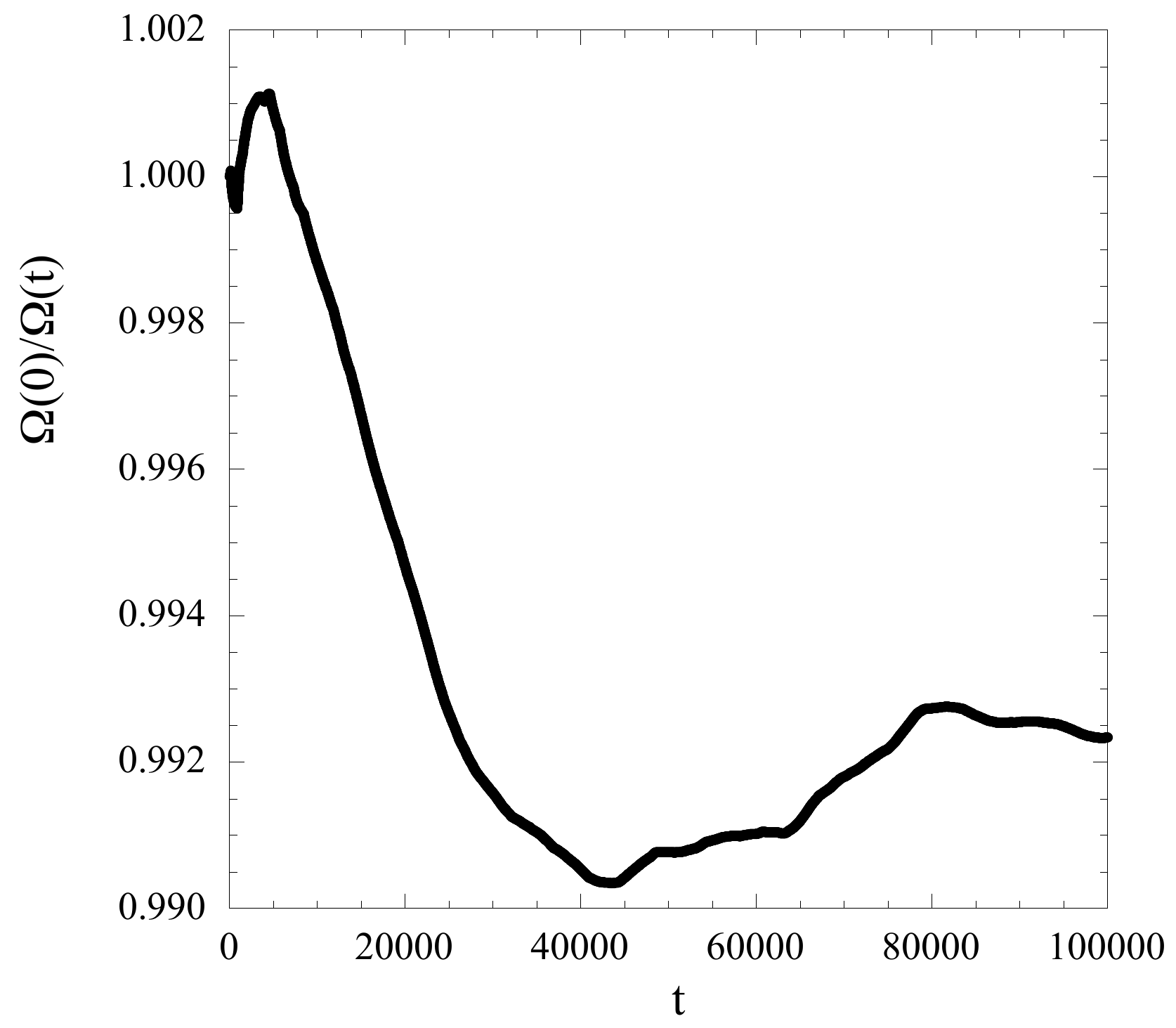}
\vspace{-0.5cm}
\caption{\label{fig:Fig12-Metric}(a) The linear time-dependence of the inverse  wealth metric indicates that the system is effectively ergodic for $\lam=0.5$ (upper line) and $\lam=0.9$ (bottom line). The inverse slopes are $3777$ and $44300$ for $\lam=0.5$ and $\lam=0.9$, respectively. (b) The inverse wealth metric for $\lam=1.0$ does not increase linearly and the system is not ergodic ($N=5000$, $f=0.01$, $\mu=0.1$).}
\end{figure}

The linear time-dependence of $\Omega(0)/\Omega(t)$ shown in Fig.~\ref{fig:Fig12-Metric}(a) for $\lam < 1$ implies that the system is effectively ergodic for $\lam < 1$.
In contrast, $\Omega(0)/\Omega(t)$ for $\lam = 1$ does not increase linearly with $t$ [see Fig.~\ref{fig:Fig12-Metric}(b)], and hence the system is not ergodic. Similar results are found for $\lam > 1$.

\subsection{\label{sec:mobility}Economic mobility}

In a system with economic mobility, poorer agents can become wealthier and richer agents can become poorer. In contrast, agents in a system with very low economic mobility rarely change their rank and the very rich become richer~\cite{Cardoso}.

\begin{figure}[tbp] 
\centering
\includegraphics[scale=0.50]{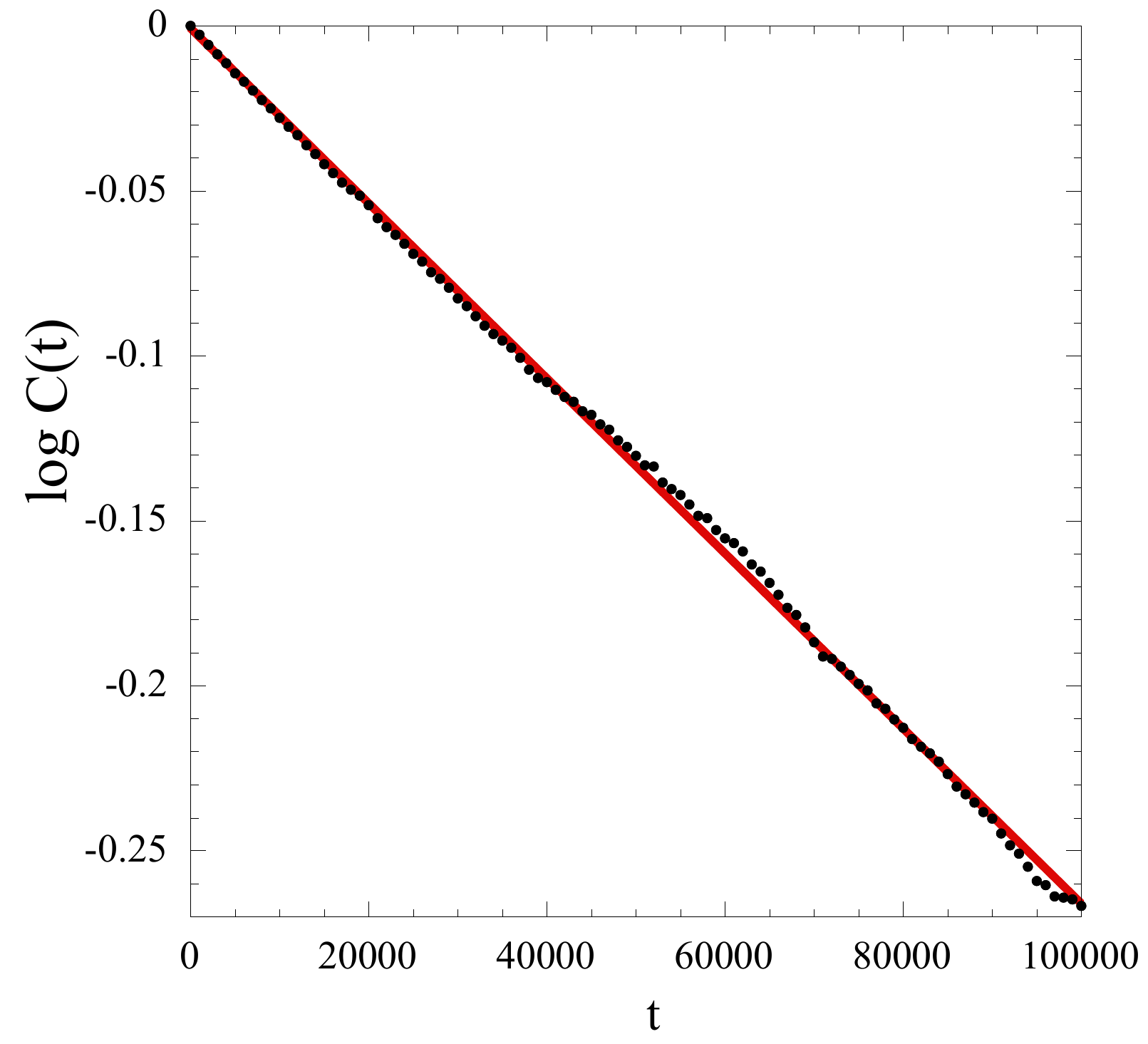}
\includegraphics[scale=0.49]{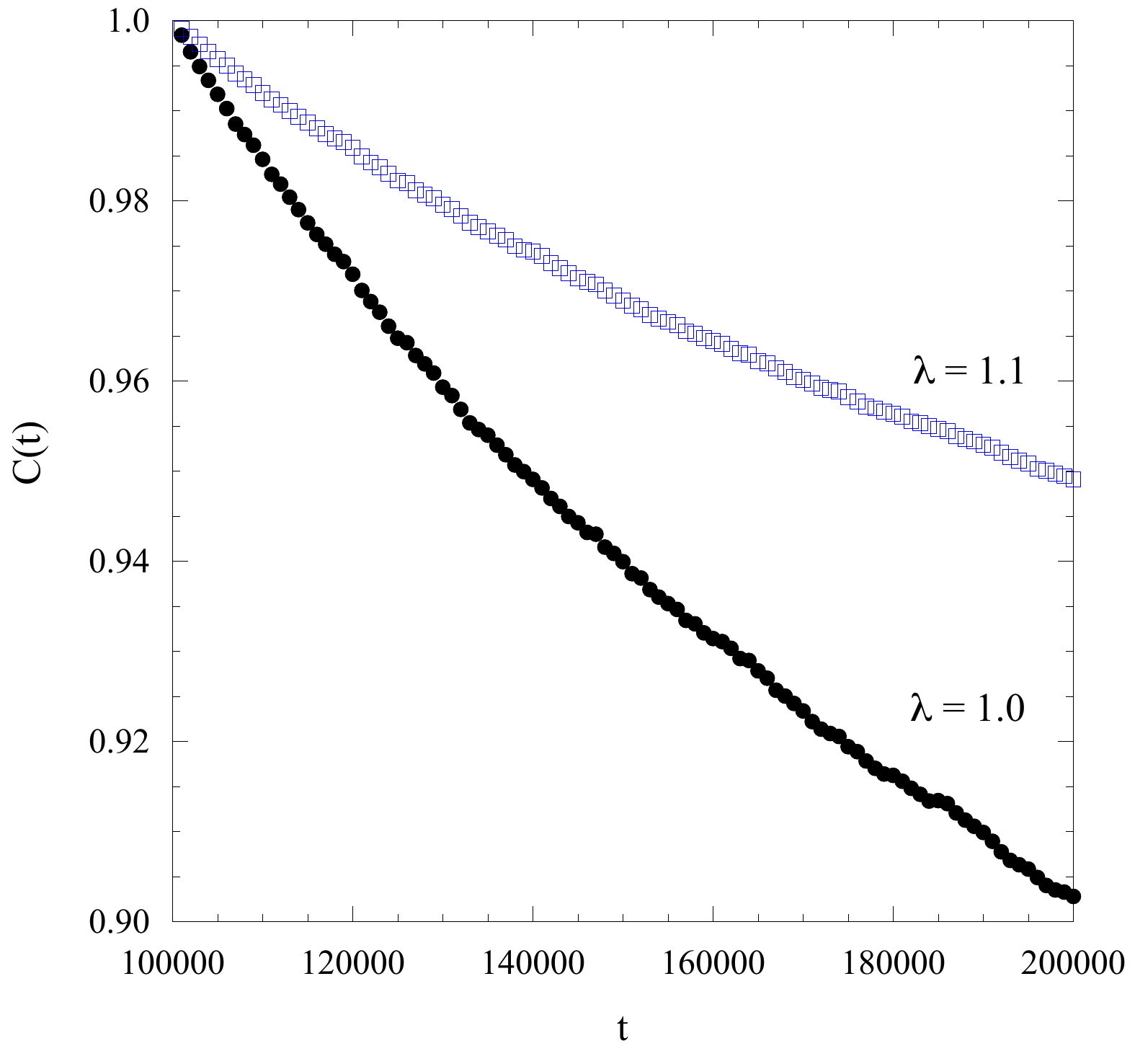}
\vspace{-0.5cm}
\caption{\label{fig:Fig11-Correlation}(a) The rank correlation decays exponentially (red line) for $\lam=0.9$, indicating that the mobility is nonzero. (b) The Pearson correlation function for $\lam=1.0$ ($\bullet$) and $\lam=1.1$ ($\square$). For $\lam \geq 1$, $C(t)$ remains nonzero even as $t\rightarrow \infty$, which indicates that the rank of an agent remains correlated and there is no mobility ($N=2500$, $f=0.1$,  $\mu=0.001$).}
\end{figure}

To determine the mobility, we rank the agents according to their wealth at various times and compute the correlation function $C(t)$ of the ranks of the agents once a steady state has been reached for $\lam < 1$. The rank correlation function $C_i(t)$ of agent $i$ is defined as
\be
C_i(t) = \frac{\lb R_i(t) R_i(0) \rb - \lb R_i \rb^2} {\lb R_i^2\rb - \lb R_i \rb^2},\label{pear-cor}
\ee
where $R_{i}(t)$ is the rank of agent $i$ at time $t$ and $\lb R_i \rb=N/2$. 
The corresponding quantity for $\lam \geq 1$, where a steady state is not reached, is the Pearson correlation function given by~\cite{pearson}
\be
\label{pear-cor2}
C_i(t) = \frac{\big [ R_{i}(t) - \lb R(t)\rb\big]\big [ R_{i}(0) - \lb R(0)\rb\big ]} {\sqrt{\big [\big (R_{i}(t) - \lb R_{i}(t)\rb\big )^{2}\big ]\big [\big (R_{i}(0) - \lb R_{i}(0)\rb\big )^{2}\big ]}},
\ee
The correlation function averaged over all agents is $C(t) = (1/N) \sum_i C_i(t)$. As can be seen from Fig.~\ref{fig:Fig11-Correlation}(a), $C(t)\rightarrow 0$ as $t\rightarrow \infty$ for $\lam<1$, which indicates that the rank of an agent as $t\rightarrow \infty$ is not correlated with its rank at $t = 0$ and the agents have a nonzero economic mobility for $\lam < 1$. The $\lam$ dependence of the average time that the richest agent remains the richest is discussed in Sec.~\ref{sec:times}. In contrast, in Fig.~\ref{fig:Fig11-Correlation}(b) we see that $C(t)$ approaches a constant for $\lam \geq 1$, indicating that the ranks are strongly correlated at different times, and there is no economic mobility.

\section{\label{sec:equilibrium}Equilibrium, not just steady state}

We have seen that the GED model approaches a steady state and is effectively ergodic for $\lam < 1$. In the following, we will show that a reasonable definition of the total energy yields an energy distribution that is consistent with the Boltzmann distribution. The existence of the latter is consistent with the idea that that the system is not just in a steady state, but is in thermodynamic equilibrium for $\lam < 1$.

As discussed in Ref.~\cite{kleinmf} (following paper), a \mft\ treatment of the GED model yields a quantity that can be interpreted as the total energy of the system at time $t$
\be
E(t)= \sum_{i=1}^N [1 - w_i(t)]^2. \label{eq:Etotal} \\
\ee
Note that the energy is zero if all agents have the same wealth ($w_i=1$).
Equation~\eqref{eq:Etotal} yields a mean energy that is extensive, that is, $\lb E \rb \propto N$ for fixed values of $\lam$, $f$, and $\mu$. For example, for $\lam = 0.99$, $f=0.01$, and $\mu=0.1$, we find that $\lb E_{N=4000}\rb/\lb E_{N=1000}\rb = 336.4/84.2 = 3.995 \approx 4$.

\begin{figure}[tbp] 
\includegraphics[scale=0.48]{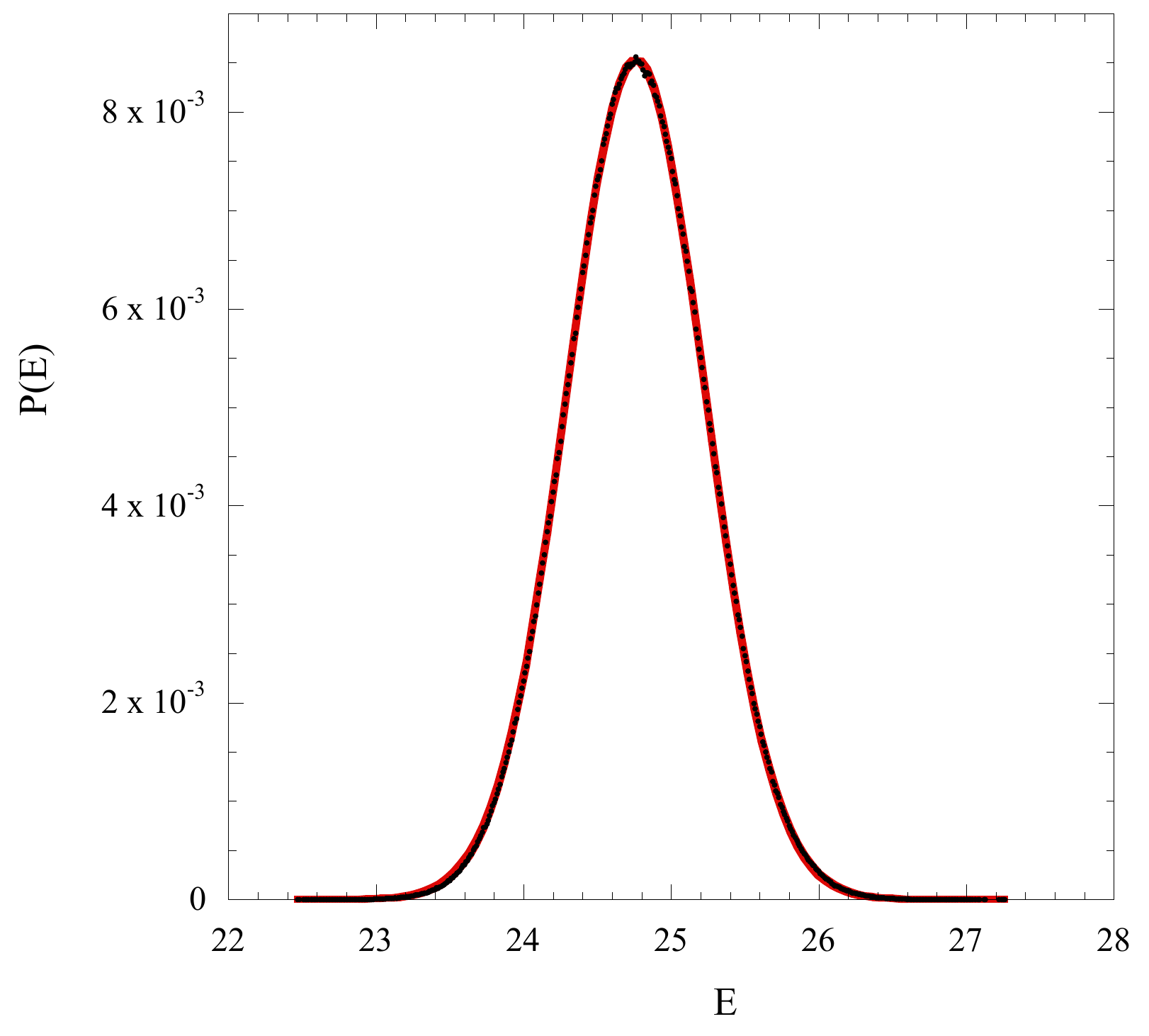}
\includegraphics[scale=0.48]{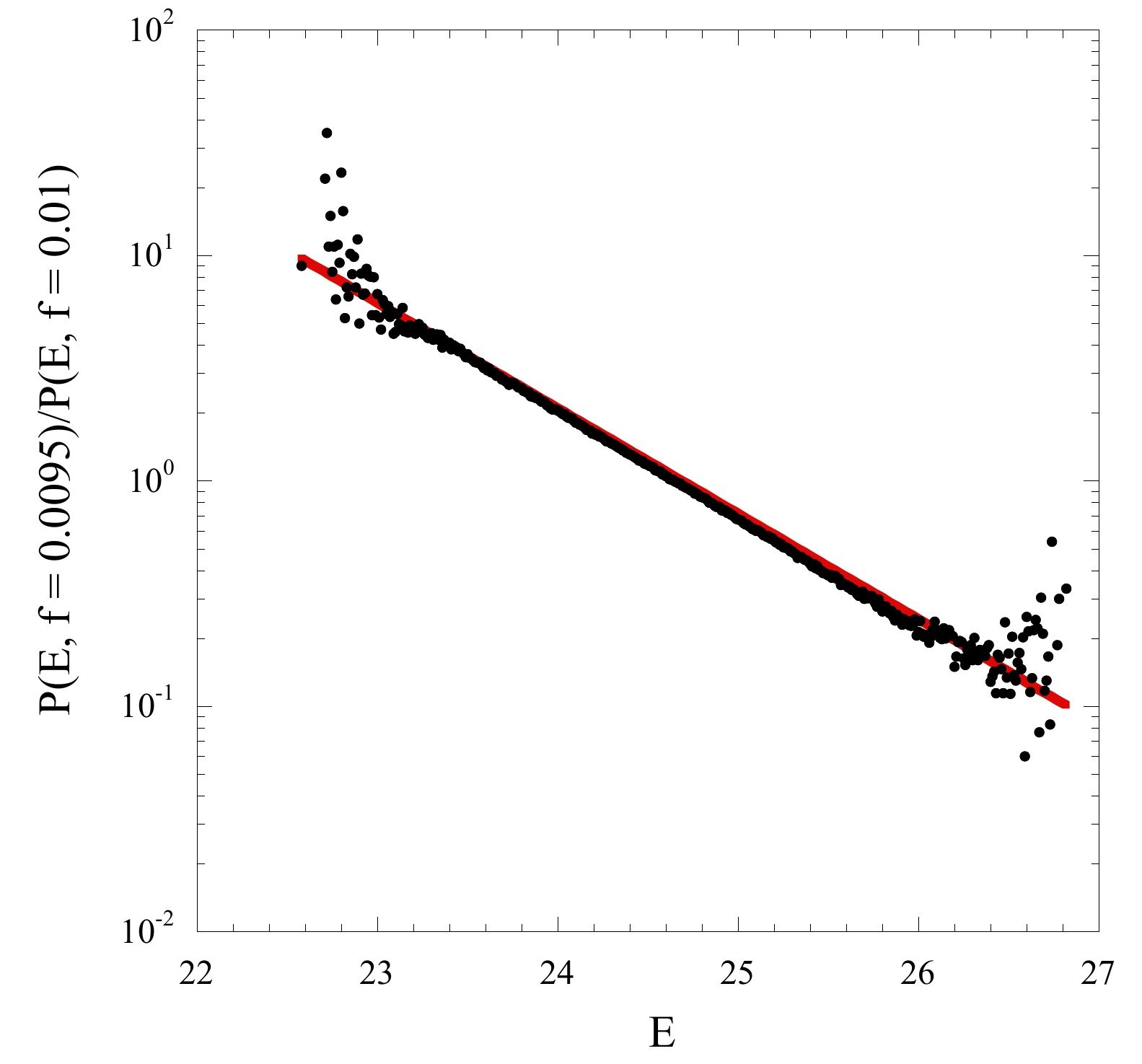}
\vspace{-0.3cm}
\caption{\label{fig:gaussian}(a) The energy probability density $P(E)$ for $N=5000$, $\lam=0.8$, $f=0.01$, and $\mu=0.1$ is consistent with the Gaussian distribution $P(E) \propto \exp(E - \overline{E})^2/\sigma_E^2)$, with $\overline{E}=24.8$ and $\sigma_E = 0.44$ [red curve]. (b) The ratio $P(E,f_1=0.01)/P(E,f_2=0.0095)$ is consistent with an exponential of the form $e^{-\Delta \beta E}$, with $\Delta \beta \approx 1.08$ (red curve).}
\end{figure}

The probability density $P(E)$ is shown in Fig.~\ref{fig:gaussian}(a) for $N=5000$, $\lam=0.8$, $f=0.01$, and $\mu=0.1$. As expected, $P(E)$ is fit well by a Gaussian. Fits of $P(E)$ to a Gaussian become less robust as $\lam \to 1$ for fixed $N$, $f$, and $\mu$. We will discuss the behavior of $P(E)$ for larger values of $\lam$ in Sec.~\ref{sec:mft}.

If the system is in thermodynamic equilibrium, we expect that the probability density $P(E,\beta)$ to be proportional to $g(E) e^{-\beta E}$, where $\beta$ is an effective inverse temperature that depends on   $\lam$, $f$, and $\mu$, and $g(E)$ is the density of states, which is independent of $\lam$, $f$, and $\mu$ and hence independent of $\beta$. Because $g(E)$ is independent of $\beta$, the ratio $P(E,\beta_1)/P(E,\beta_2)$ is an exponential proportional to $ \exp[-(\beta_1 - \beta_2)]E$ if the system is characterized by the Boltzmann distribution with the energy given by Eq.~\eqref{eq:Etotal}. The range of values of $E$ over which this ratio is nonzero and finite is limited by the overlap of the two probabilities, which becomes smaller as $N$ is increased. 
In Fig.~\ref{fig:gaussian}(b) we see that the ratio $P(E,\beta_2)/P(E,\beta_1)$ is consistent with the Boltzmann distribution $e^{-(\beta_2-\beta_1)E}$ for $N=5000$, $\lam=0.8$, and $\mu=0.1$, with $\beta \propto f^{-1}$, $f_2=0.0095$ and $f_1=0.01$; a larger value of $f$ (for fixed values of $\lam$ and $\mu$) corresponds to a smaller value of $\beta$ and hence a higher value of the effective temperature. The dependence of the effective temperature $\beta^{-1}$ on $f$ is consistent with the association of $f$ with the presence of both additive and multiplicative noise in the system~\cite{kleinmf}. Similar results hold for two similar values of $\lam$ for fixed $f$ and $\mu$. The existence of an energy, and the observation that its probability is proportional to the Boltzmann distribution implies that the system can be considered to be in thermodynamic equilibrium, at least for $\lam = 0.8$.

\section{\label{sec:phasetransition}Characterization of the phase transition}

\subsection{Fixed number of agents}\label{sec:fixedagents}

The numerical results discussed in this section are for $N=5000$, $f=0.01$, and $\mu=0.1$. 
Averages are taken over a time of $10^6$ ($N \times 10^6$ exchanges) after a transient time of $10^6$. 
The major source of uncertainty in our estimations of the values of the various critical exponents is the choice of the range of values to be retained in the least squares fits.

Because we will characterize the approach to the phase transition at $\lam=1$ in terms of power laws, it is convenient to define the quantity
\be
\e \equiv 1 - \lam,
\ee
and will assume that $\e > 0$.
 
To characterize the phase transition at $\e=0$, we need to identify an order parameter. A common measure of income or wealth inequality is the Gini coefficient $G_n$~\cite{gini}. If all agents have the same wealth, $G_n=0$. In contrast, if one agent has all the wealth, $G_n=1$, corresponding to the maximum degree of inequality. These characteristics appear to make $1-G_n$ a reasonable choice of the order parameter. However, because the wealth distribution reaches a steady state for $\lam < 1$, the fluctuations of the Gini coefficient are zero in the limit $N \to \infty$, and hence the susceptibility, which would be associated with the variance of $G_n$, would be zero, making $G_n$ an inappropriate choice of the order parameter.

\begin{figure}[t]
\includegraphics[scale=0.55]{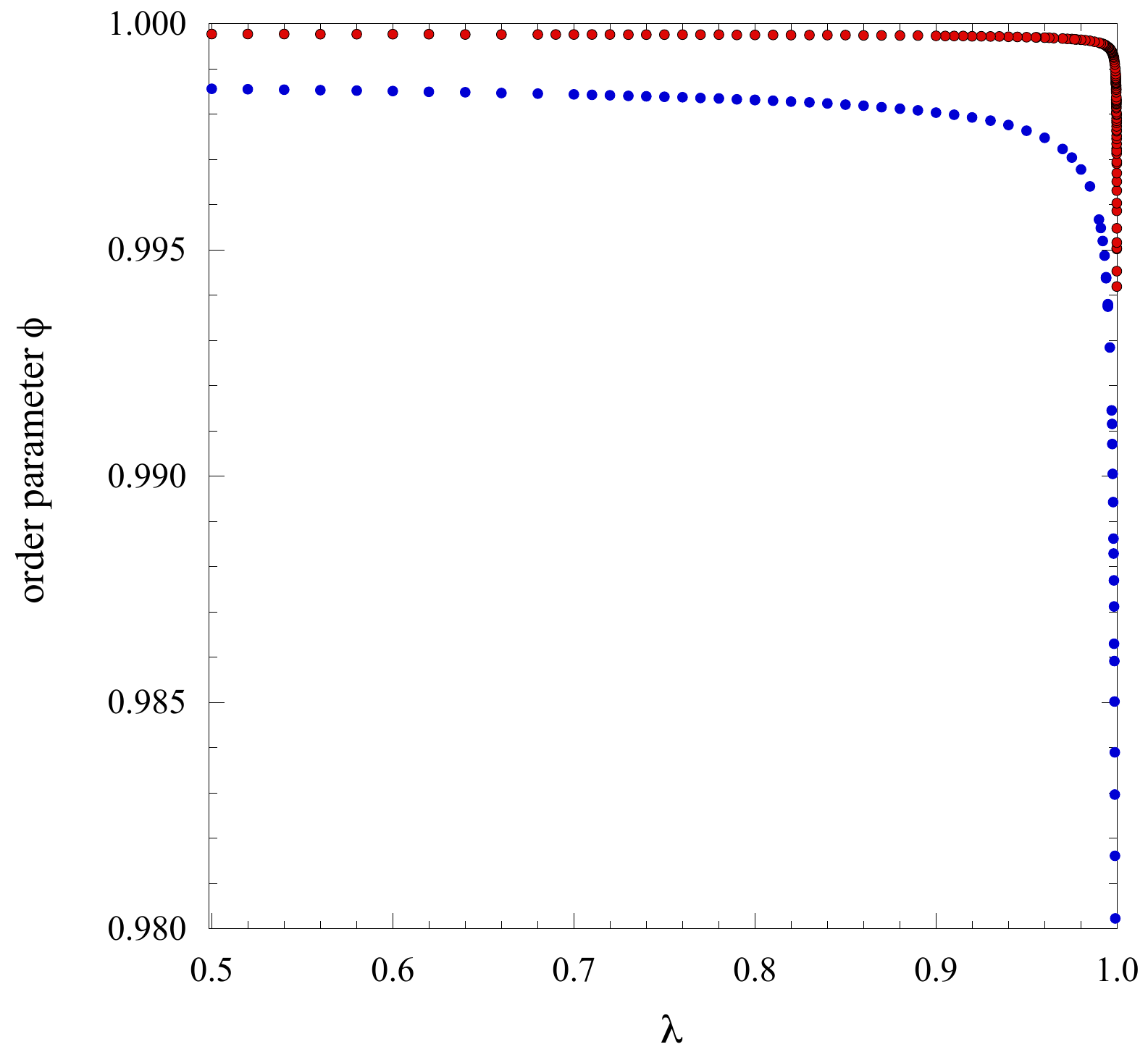}
\vspace{-0.25cm}
\caption{\label{fig:phasediagram}The $\lam$-dependence of the order parameter $\phi$, the fraction of wealth held by all agents except the richest agent, for $N=5000$ (top curve) and $N=1000$ (bottom curve) with $f=0.01$ and  $\mu=0.1$. In the limit $N \to \infty$ the order parameter becomes a step function with $\phi = 1$ for $\lam < 1$ and $\phi = 0$ for $\lam \geq 1$.}
\end{figure}

{\it The order parameter and the value of $\beta$}. Another  choice of the order parameter is the fraction of the wealth held by all the agents except the richest agent, that is,
\be
\phi = \frac{N - \wmax}{N}, \label{eq:phi}
\ee
where $\wmax$ is the wealth of the richest agent. For $\lam < 1$ we find that $\wmax \ll N$ and depends  weakly on $N$ for fixed values of $\lam$, $f$, and $\mu$. For example, $\wmax = 2.0$ for $N=1000$ and $\wmax = 2.1$ for $N=4000$, with $\lam=0.99$, $f=0.01$, and $\mu=0.1$.  The weak dependence of $\wmax$ on $N$ implies that $\phi$ defined in Eq.~\eqref{eq:phi} approaches one as $N \to \infty$ for $\lam < 1$ independently of the value of $\lam$ (see Fig.~\ref{fig:phasediagram}). Hence, the value of the critical exponent $\beta$ associated with the order parameter is $\beta = 0$. We also point out that for $\lambda \geq 1$, the order parameter is zero in the limit  $N \to \infty$. If this branch is continued to $\lambda < 1$, $\phi$ will remain zero (because all agents except the richest have zero wealth), indicating that there is hysteresis. This behavior of $\phi$
raises the question of the order of the transition at $\lambda = 1$. One possibility is that the transition  is a spinodal~\cite{bigklein}.

\begin{figure}[t] 
\includegraphics[scale=0.48]{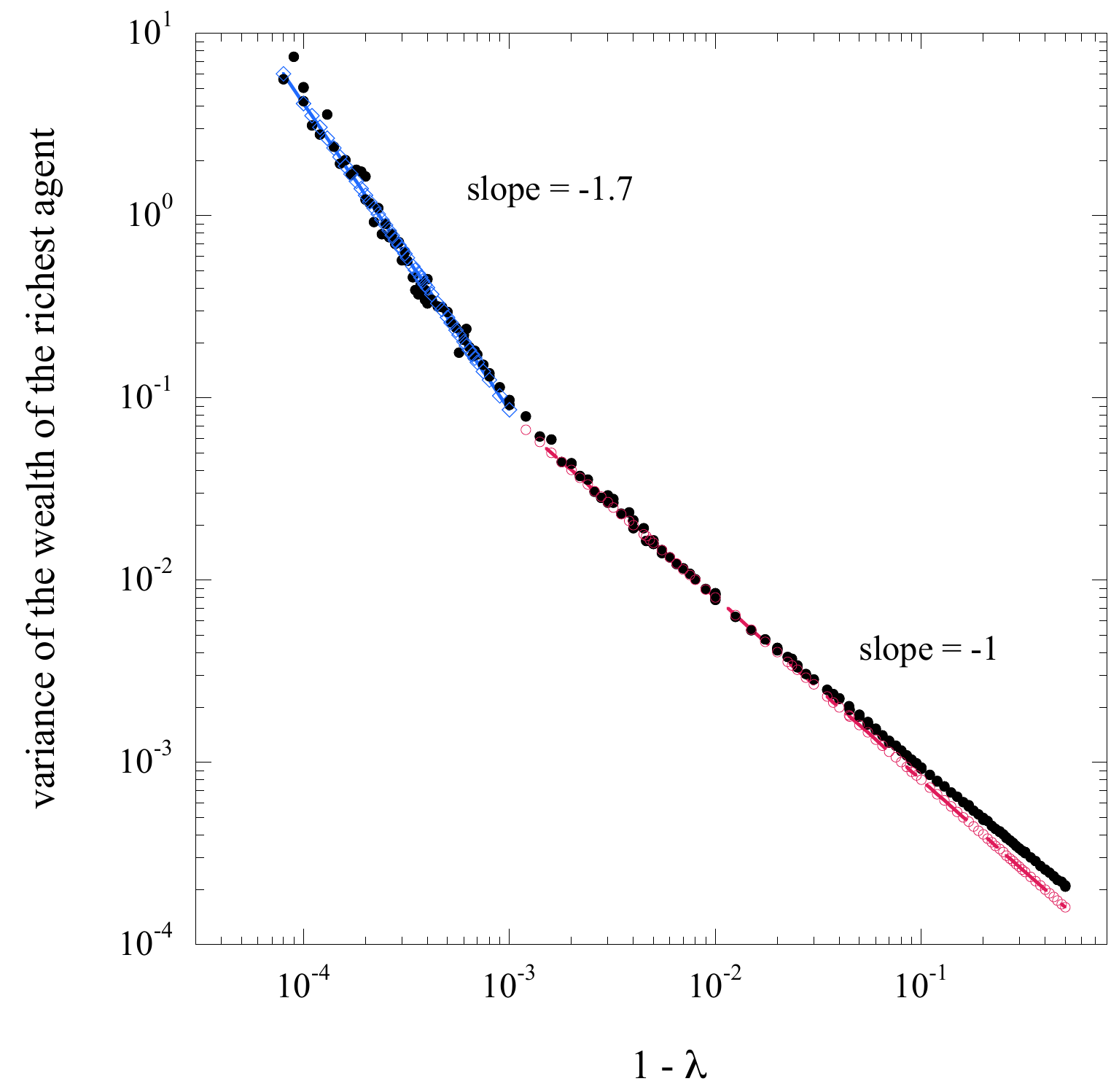}
\includegraphics[scale=0.48]{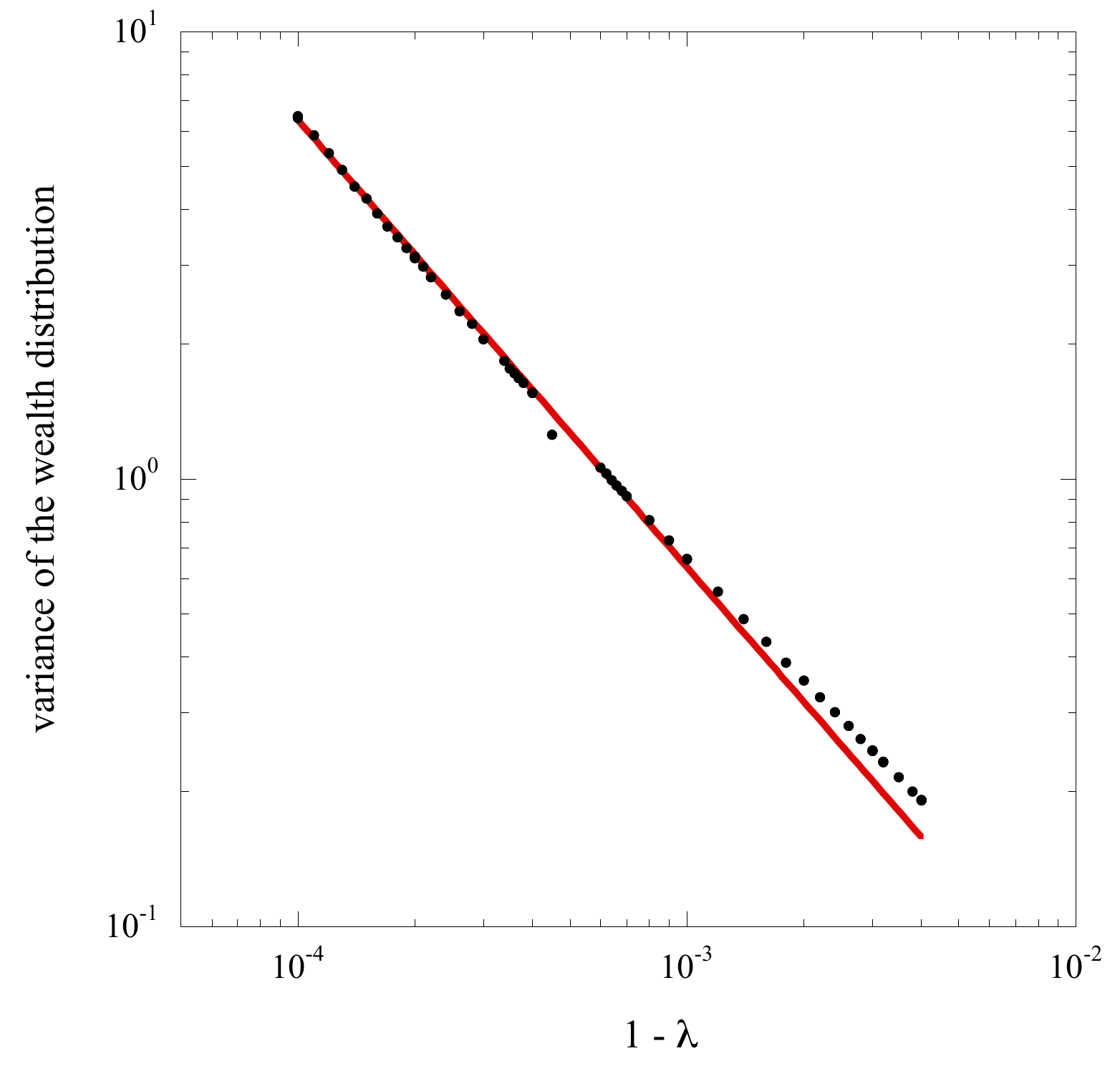}
\vspace{-0.2in}
\caption{\label{fig:chivar}(a) The $\e$-dependence of the variance of the wealth of the richest agent, shows crossover behavior. Fits of the variance to a power law for $\e=(1-\lam) \geq 2 \times 10^{-3}$ give an effective exponent close to one (red curve); fits for smaller values of $\e$ give an effective exponent equal to $\approx 1.7$ (blue curve). (b) The $\e$-dependence of $C_2$, the variance of the wealth of the individual agents, shows less curvature, and gives an effective exponent close to 1.0 (red curve) if fits are made for $\e < 0.0015$; the effective exponent is in the range $[0.93, 1.0]$ depending on the values of $\e$ that are included in the least squares fits ($N=5000$, $f=0.01$, $\mu = 0.1$).}
\end{figure}

{\it The susceptibility and the value of $\gamma$}. The $\e$-dependence of the variance of $\wmax$ is shown in Fig.~\ref{fig:chivar}(a) and can be fit to a power law to give an effective exponent for the susceptibility close to one for $\e \geq 2 \times 10^{-3}$; fits for smaller values of $\e$ give an effective exponent approximately equal to 1.7. Both estimates indicate that the variance of the wealth of the richest agent diverges strongly, and hence we choose the order parameter to be as defined in Eq.~\eqref{eq:phi}.

More consistent results for the susceptibility can be found from $C_2$, the variance of the wealth of a single  agent averaged over all agents, and not just the variance of the wealth of the richest agent. In Fig.~\ref{fig:chivar}(b) we see that there is less curvature in the plot of $\log C_2$ versus $\log \e$, and we find an effective exponent close to one if fits are made for $\e < 0.0015$. If values of $C_2$ are included for larger values of $\e$, the effective exponent from the least squares fits is in the range $[0.93, 1.0]$. Given these much better fits, we associate the susceptibility with $NC_2$:
\be
\chi = N C_2, \label{eq:chi}
\ee
and conclude that the critical exponent associated with the susceptibility is $\gamma \approx 1$.

\begin{figure}[t] 
\includegraphics[scale=0.51]{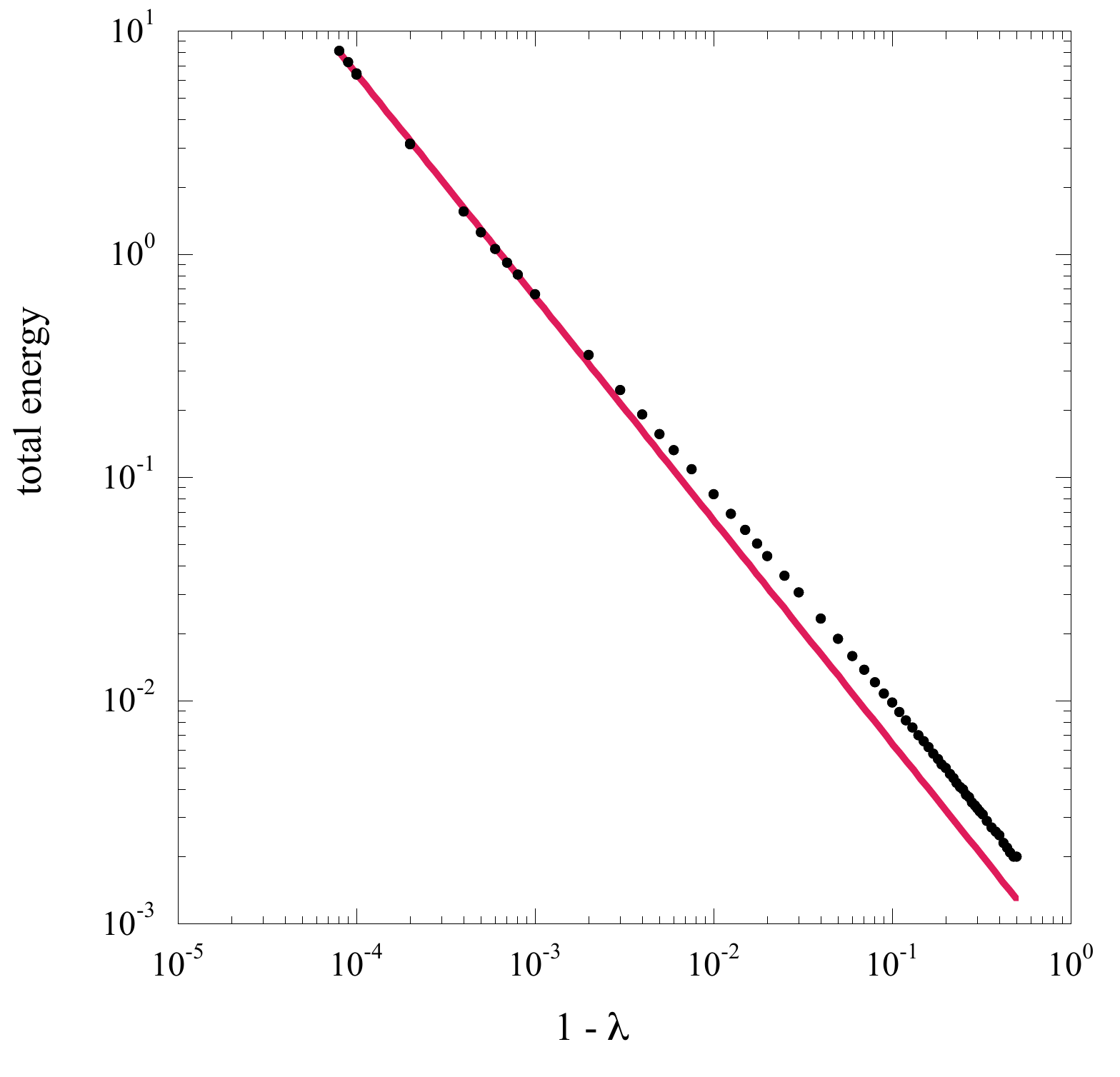}
\includegraphics[scale=0.51]{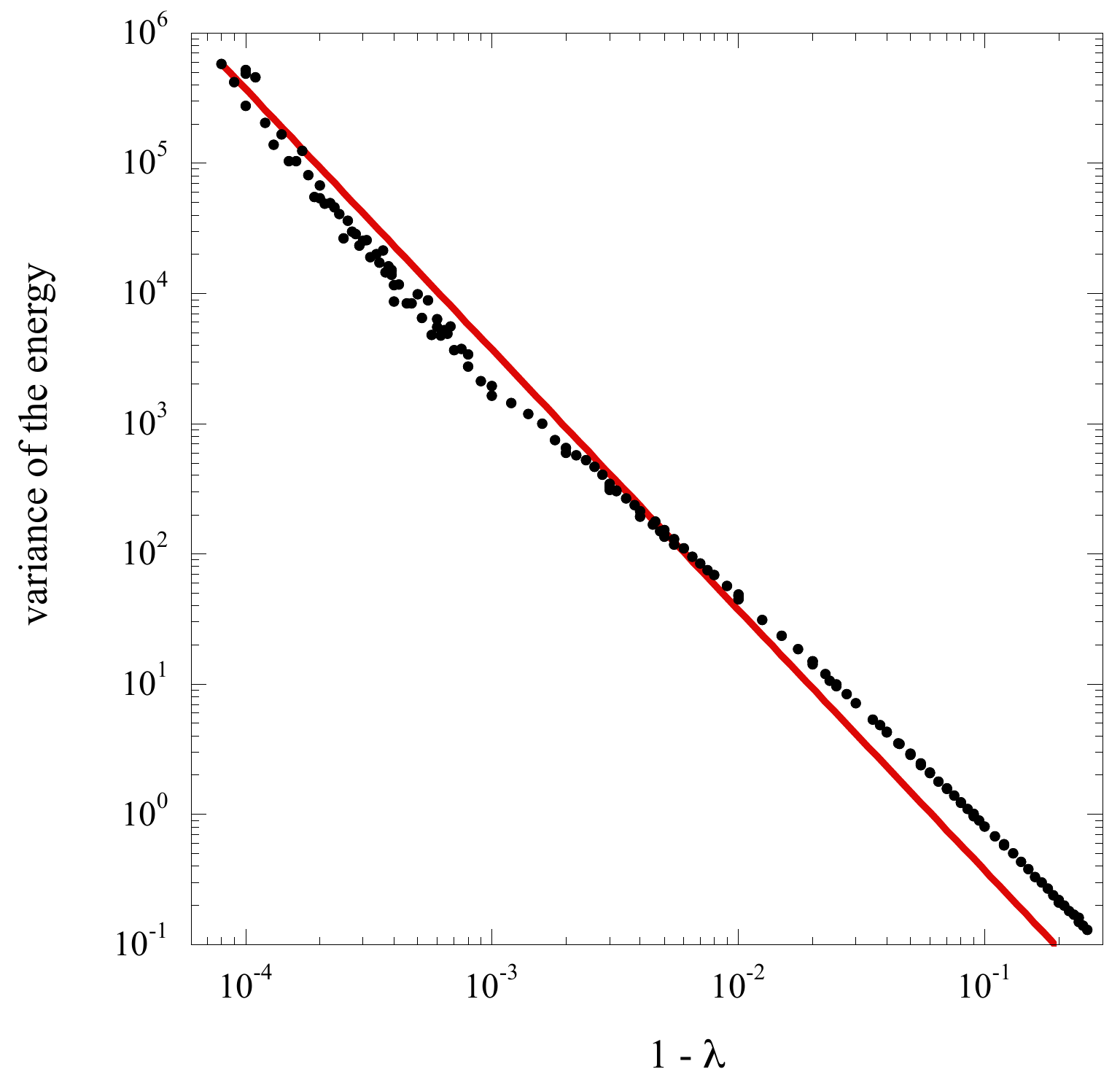}
\vspace{-0.3cm}
\caption{\label{fig:energy}(a) The divergent $\e$-dependence of the mean total energy $\ebar$ for fixed $N$, $f$, and $\mu$ is consistent with the power law $\e^{-1}$ (red line). Least squares fit of $\ebar$ yield values of the divergence in the range $[0.93, 1.03]$. (b) The $\e$-dependence of the heat capacity $C$ is consistent with the power law $\e^{-2}$ (red line). 
Least squares fits of $C$ yield values of the effective exponent in the range $[1.8, 2.1]$ ($N=5000$, $f=0.01$, $\mu=0.1$).}
\end{figure}

{\it The mean energy, heat capacity, and the value of $\alpha$}. Although the total rescaled wealth is a constant, the energy depends on the way the wealth is distributed. We use Eq.~\eqref{eq:Etotal} to determine the mean energy by averaging the quantity $\sum_{i=1}^N(1-w_i)^2$ over many realizations. Our results for the mean energy $\ebar$ are shown in Fig.~\ref{fig:energy}(a). We see that the $\e$-dependence of $\ebar$ is consistent with $\e^{1 - \alpha}$ as $\e \to 0$ with $\alpha \approx 2$. This divergence of $\ebar$ is inconsistent with equilibrium statistical mechanics, which requires that the total energy be finite for finite values of $N$.

We define the heat capacity $C$ to be the variance of the total energy defined in Eq.~\eqref{eq:Etotal}. The $\e$-dependence of of $C$ is consistent with the dependence $\e^{-\alpha}$ with the exponent $\alpha = 2$ as shown in Fig.~\ref{fig:energy}(b). Our determination of the value of $\alpha$ depends on the range of values of $\e$ that are included in the least squares fits and are in the range $[1.8, 2.1]$.

In summary, our numerical results for the critical exponents $\alpha$, $\beta$, and $\gamma$, determined as $\e$ is varied for a given value of $N$ are consistent with
\be
\beta = 0, \mbox{ } \gamma = 1, \mbox{ and } \alpha = 2 \qquad \mbox{(fixed $N$, $f$, and $\mu$).}
\ee
These numerical values are inconsistent with the usual scaling law
\be
\alpha + 2\beta + \gamma = 2. \label{eq:scalinglaw}
\ee
However, the value of $\alpha$ determined from the power law behavior of the heat capacity is consistent with the $\lam$-dependence of the mean total energy, that is, $C = \partial \ebar/\partial \lam$.

\subsection{\label{sec:mft}Mean-field theory and fixed Ginzburg parameter}

Although it is natural to determine the critical behavior of the GED model as the critical point is approached for a fixed number of agents, our numerical results for the critical exponents are not consistent with the scaling law, Eq.~\eqref{eq:scalinglaw}, nor consistent with equilibrium statistical mechanics because the mean energy per agent diverges as the critical point is approached even for finite values of $N$ and fixed values of $f$ and $\mu$. Simulations also show that $\ebar$ and the heat capacity $C$ are proportional to $N$ for fixed values of $\lam$, $f$, and $\mu$ so that the divergent behavior of $\ebar$ is not removed by first taking the 
limit $N \to \infty$ before taking the limit $\e \to 0$.

In Ref.~\cite{kleinmf} (following paper) a mean-field treatment of the GED model is developed based on the random exchange of wealth between an agent chosen at random and an agent whose wealth is assigned to be equal to the mean wealth of the remaining agents. The \mf\ treatment predicts that the critical exponents are given by
\be
\beta=0, \mbox{ } \gamma = 1, \mbox{ and } \alpha=1. \label{eq:mftpredictions}
\ee
The predicted \mf\ values of the exponents in Eq.~\eqref{eq:mftpredictions} are consistent with Eq.~\eqref{eq:scalinglaw}. The \mf\ theory~\cite{kleinmf} also predicts that the mean energy per agent approaches a constant as $\e \to 0$.

To compare the mean-field predictions with the simulations for different values of $N$ we need to account for the fact 
$f$ and $\mu$ are rates and depend on the definition of the unit of time.
It is shown 
in the \mft\ of Ref.~\cite{kleinmf} that to achieve a consistent thermodynamic description of the GED model, we need to rescale $f$ and $\mu$ as we change $N$ so that
\be
f = f_0/N \mbox{ and } \mu = \mu_0/N. \label{eq:scalefandmu}
\ee

Another condition for the applicability of the \mf\ treatment of the GED model is that the Ginzburg parameter $G$, defined as
\be
G \equiv \frac{\mu_0 N (1 - \lam)}{f_0^2}, \label{eq:G}
\ee
be much greater than one and be held fixed as $\e \to 0$. As has been found for the long-range and fully connected Ising models~\cite{bigklein, lou, kangspecificheat}, 
we will find that the \mft\ predictions for the critical behavior of the energy and heat capacity of the GED model are consistent with equilibrium statistical mechanics only if the Ginzburg parameter is held fixed as the critical point is approached. We will also see  that the results for $\beta$ and $\gamma$ do not depend on keeping $G$ fixed as has been found for other fully connected models~\cite{bigklein, lou, kangspecificheat}. 

We emphasize that if $\mu$ and $f$ are not rescaled in the simulations, the energy per agent would diverge as $N\rightarrow \infty$ even for fixed Ginzburg parameter.

To compare the \mft\ predictions to the simulations, we choose $N_0 = 5000$, $f_0 = 0.01$, and $\mu_0 = 0.1$, with $f= (N_0/N)f_0$ and $\mu = (N_0/N)\mu_0$. For a particular choice of the value of $\lam$, we determine the value of $N$ needed to keep the value of $G$ in Eq.~\eqref{eq:G} fixed at $G=10^6$. Our simulations are for $0.80 \leq \lam \leq 0.998$ and $5 \times 10^3 \leq N \leq 5 \times 10^5$.

\begin{figure}[t] 
\includegraphics[scale=0.5]{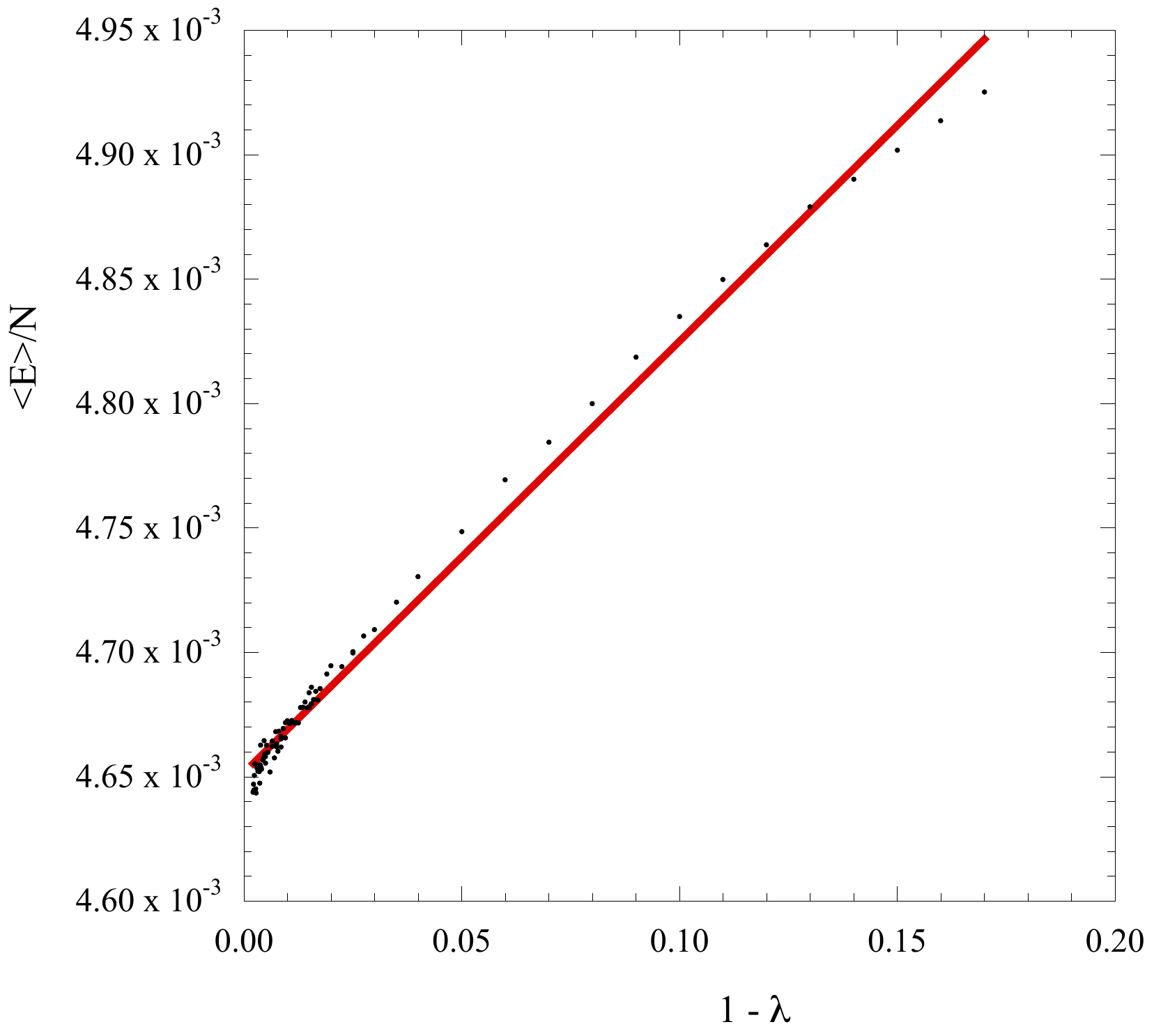}
\vspace{-0.2in}
\caption{\label{fig:enG10}The $\e$ dependence of $\ebar/N$, the mean energy per agent, for $G=10^6$ is consistent with the linear $\e$-dependence $a_0 + a_1\e$ as $\e \to 0$, with $a_0 \approx 0.005$ and $a_1 \approx 0.002$ (red line). The $\e$ dependence of $\ebar/N$ is given by Eq.~\eqref{eq:egeneral}.}
\end{figure}

Because our results for $\ebar$ and $C$ depend on keeping $G$ fixed, we first discuss their $\e$ dependence. In Fig.~\ref{fig:enG10} we see that $\ebar/N$ approaches a constant as $\e \to 0$, in contrast to its divergent $\e$ behavior for fixed $N$, $f$, and $\mu$. Because $\ebar \sim \e^{1-\alpha}$, we find that simulations for fixed Ginzburg parameter are consistent with $\alpha = 1$.

Our numerical results for $\ebar$ are consistent with the relation
\be
\ebar \sim \frac{N}{G}. \label{eq:egeneral}
\ee
Equation~\eqref{eq:egeneral} is predicted by \mft~\cite{kleinmf} for fixed $G$. For fixed $G$, Eq.~\eqref{eq:egeneral}  implies that $\ebar/N$ approaches a constant as $\e \to 0$, consistent with our simulations. In contrast,  for fixed $N$, $f$, and $\mu$ we find
\be
\ebar \sim \frac{N}{G} = 
\dfrac{Nf_0^2}{N\mu_0 \e} = \dfrac{N^2f^2}{N\mu \e} = \dfrac{Nf^2}{\mu \e} \qquad \mbox{(fixed $N$, $f$, and $\mu$)}, \label{eq:EfixedN}
\ee
where $f_0 = N f$ and $\mu_0 = N \mu$. Equation~\eqref{eq:EfixedN} implies that for fixed values of $\lam$, $f$, and $\mu$, $\ebar$ is proportional to $N$ and diverges as $\e^{-1}$ for fixed values of $N$, $f$, and $\mu$; both behaviors are consistent with our simulations.

\begin{figure}[tbp] 
\includegraphics[scale=0.6]{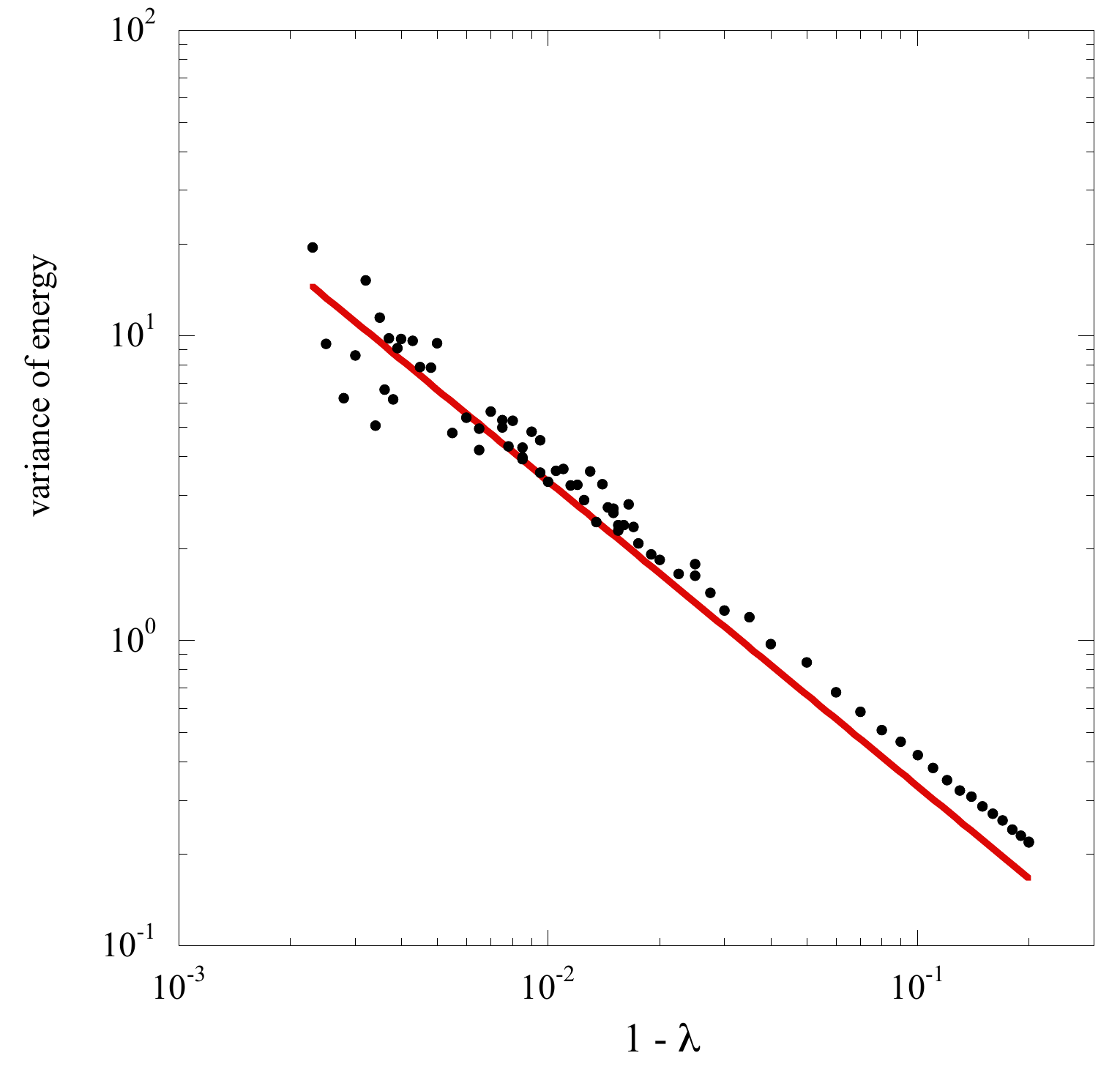}
\vspace{-0.2in}
\caption{\label{fig:specificG1}The $\e$-dependence of $C$, the variance of the total energy, for fixed Ginzburg parameter is consistent with the power law $\e^{-\alpha}$, with $\alpha = 1$ (red line). A least square fit gives an exponent of $\approx 0.91$.}
\end{figure}

The $\e$-dependence of $C$, the variance of the total energy, for fixed $G$ is shown in Fig.~\ref{fig:specificG1}. We see that $C \sim \e^{-\alpha}$, with $\alpha \approx 1$, consistent with the prediction of mean-field theory~\cite{kleinmf}.

Our numerical results for the variance of the total energy are consistent with the relation
\be
C \sim \frac{N}{G^2}. \label{eq:CG}
\ee
For fixed $G$, we have $N \propto \e^{-1}$ and hence $C \sim \e^{-1}$ and $\alpha = 1$.
For fixed values of $f$ and $\mu$ we have
\be
C \sim \dfrac{N f_0^4}{N^2 \mu_0^2\e^2} = \dfrac{N^4 f^4}{N^3 \mu^2 \e^2} \sim \frac{N}{\e^2}\qquad (\mbox{fixed $f$ and $\mu$}). \label{eq:CfixedN}
\ee
In this case $\alpha=2$ and $C$ is proportional to $N$ for fixed values of $\lambda$, $f$, and $\mu$, consistent with the simulations. 

The $\e^{-1}$ dependence of $C$ for fixed $G$ is consistent with the mean-field prediction in Ref.~\cite{kleinmf}.
Note that the variance of the total energy is proportional to $N$ for fixed $\lam$, $f$ and $\mu$, but is independent of $N$ for fixed Ginzburg parameter. This seemingly inconsistent dependence on $N$ is due to the dependence of $N$ on $\e$ for fixed Ginzburg parameter. This confusion is a consequence of the fully connected nature of the GED model. (Recall that an agent can exchange wealth with equal probability with any other agent in the system.) If the mean-field limit is taken according to the prescription of Kac et al.~\cite{kac}, then the agents would be placed, for example, on a two-dimensional lattice and the range $R$ over which the agents could exchange wealth would be finite~\cite{tim}. In this case the  Ginzburg parameter would be a function of $R$ rather than $N$, and the specific heat would be the heat capacity divided by $N$.

The $\e^{-1}$ dependence of the variance of the total energy near $\e=0$ implies that the mean energy must include a logarithmic dependence on $\lam$. For example, the form, $\ebar \sim a_0 + a_1 \e + a_L/\ln \e$, implies that $C = \partial \ebar/\partial \lam$ scales as $\e^{-1}$ with a logarithmic correction. Mean-field theory is incapable of finding logarithmic corrections, and our data for $\ebar$ is not sufficiently accurate to detect the presence of logarithmic factors.

\begin{figure}[tbp] 
\includegraphics[scale=0.7]{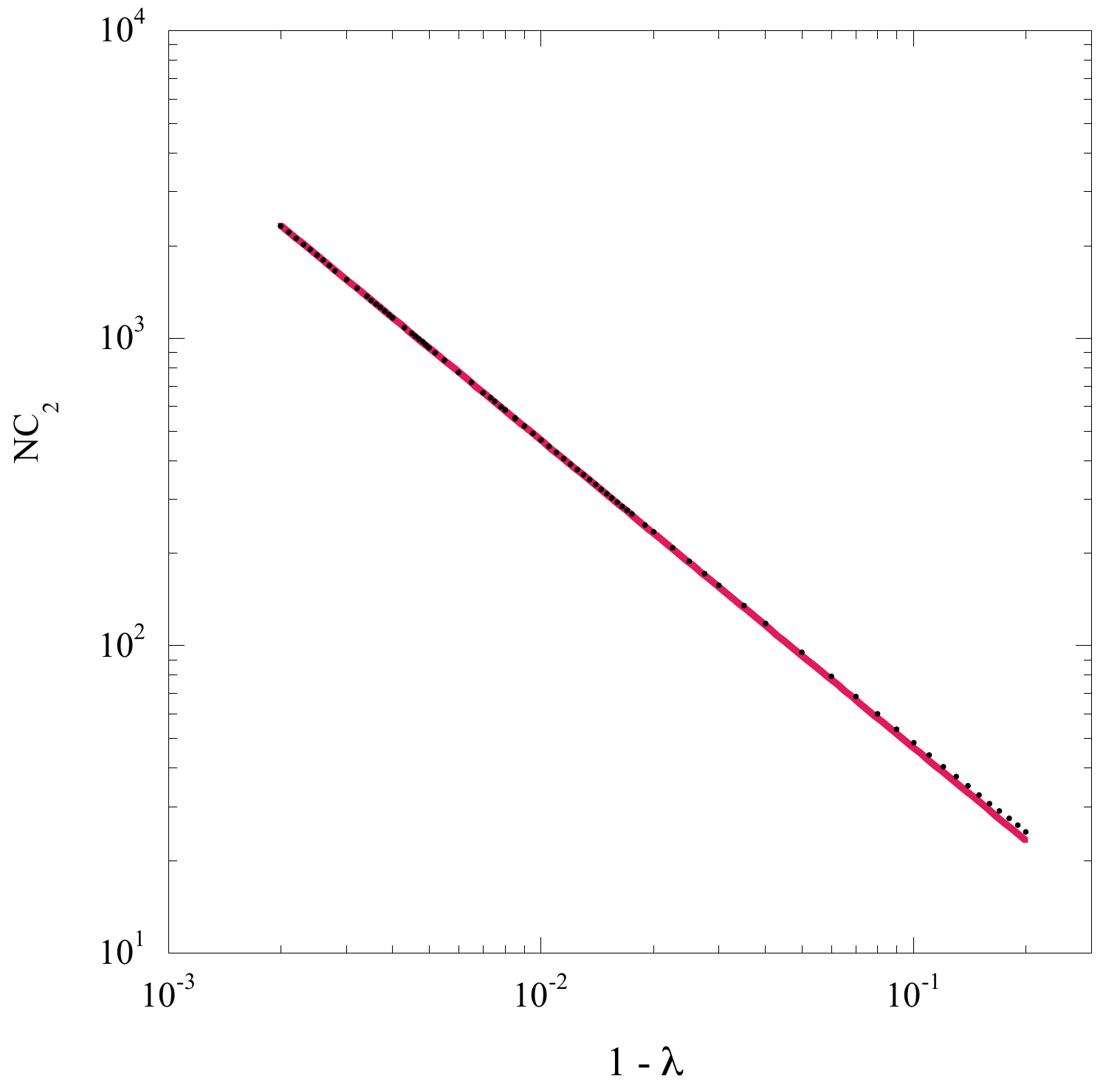}
\vspace{-0.2cm}
\caption{\label{fig:varianceG}The divergence of the susceptibility $NC_2$ is consistent with $\chi \sim \e^{-\gamma}$ with $\gamma=1$ (red line).}
\end{figure}

Although our numerical results for $\alpha$ depend on whether $N$ or $G$ is held fixed as $\lam \to 1$, our results for the order parameter and susceptibility exponents $\beta$ and $\gamma$ are independent of the nature of the approach to the phase transition. We find that $\wmax \ll N$ and is independent of $\lam$ for fixed $G$. Hence $\phi=1$ for $\lam < 1$, implying that $\beta = 0$ as was found for fixed $N$. The divergence of the susceptibility $\chi = N C_2$ shown in Fig.~\ref{fig:varianceG}(b) is consistent with $\chi \sim \e^{-\gamma}$ with $\gamma = 1$.

\section{\label{sec:times}Critical slowing down}

We find that various time scales increase rapidly as $\e \to 0$, which limits how close the simulations can approach the transition. One time scale of interest is $\trich$, the mean lifetime of the richest agent. We expect that if the mobility of the agents is nonzero, then the richest agent at $t=0$ will no longer be the richest after some time has elapsed. We define $\trich$ as the mean time that a particular agent remains the richest and assume that $\trich$ is a simple measure of the decorrelation time of the wealth of individual agents.

Another time scale of interest is the mixing time $\tau_m$ associated with the time-dependence of the wealth metric; $\tau_m$ is related to the inverse slope of the wealth metric as
\be
\label{eq:taum}
\Omega(0)/\Omega(t) = t/\tau_m. 
\ee

We also computed the time-displaced energy autocorrelation function given by
\be
C_E(t) = \frac{\lb E(t)E(0) \rb - \lb E \rb^2}{\lb E^2 \rb - \lb E \rb^2},
\ee
where $E(t)$ is the value of the energy of the system at time $t$. We find that $C_E(t)$ relaxes exponentially, and hence we can extract the energy decorrelation time $\tau_E$. Our simulation results for these various times are for fixed Ginzburg parameter ($G=10^6$).

\begin{figure}[t] 
\includegraphics[scale=0.51]{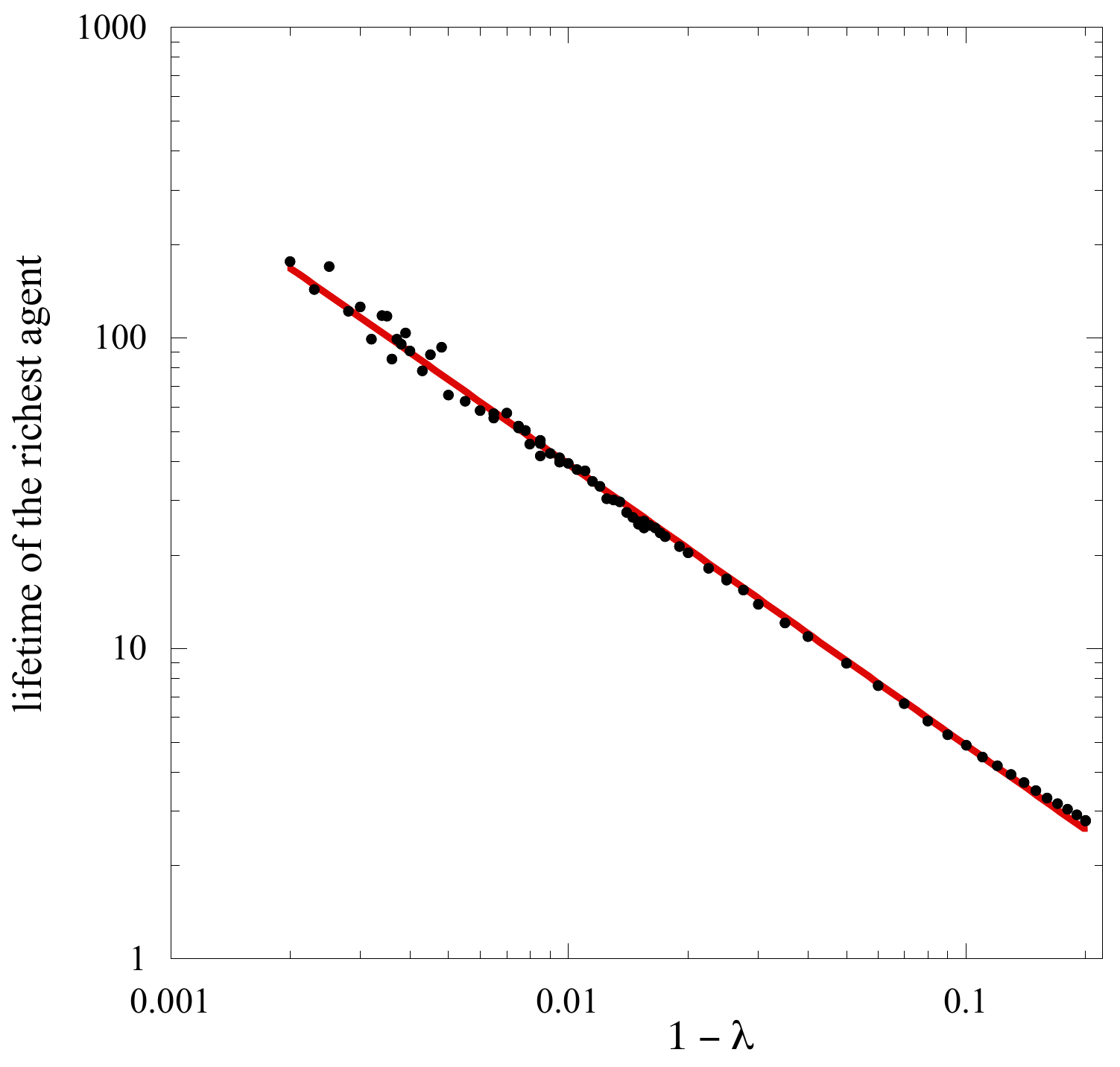}
\includegraphics[scale=0.51]{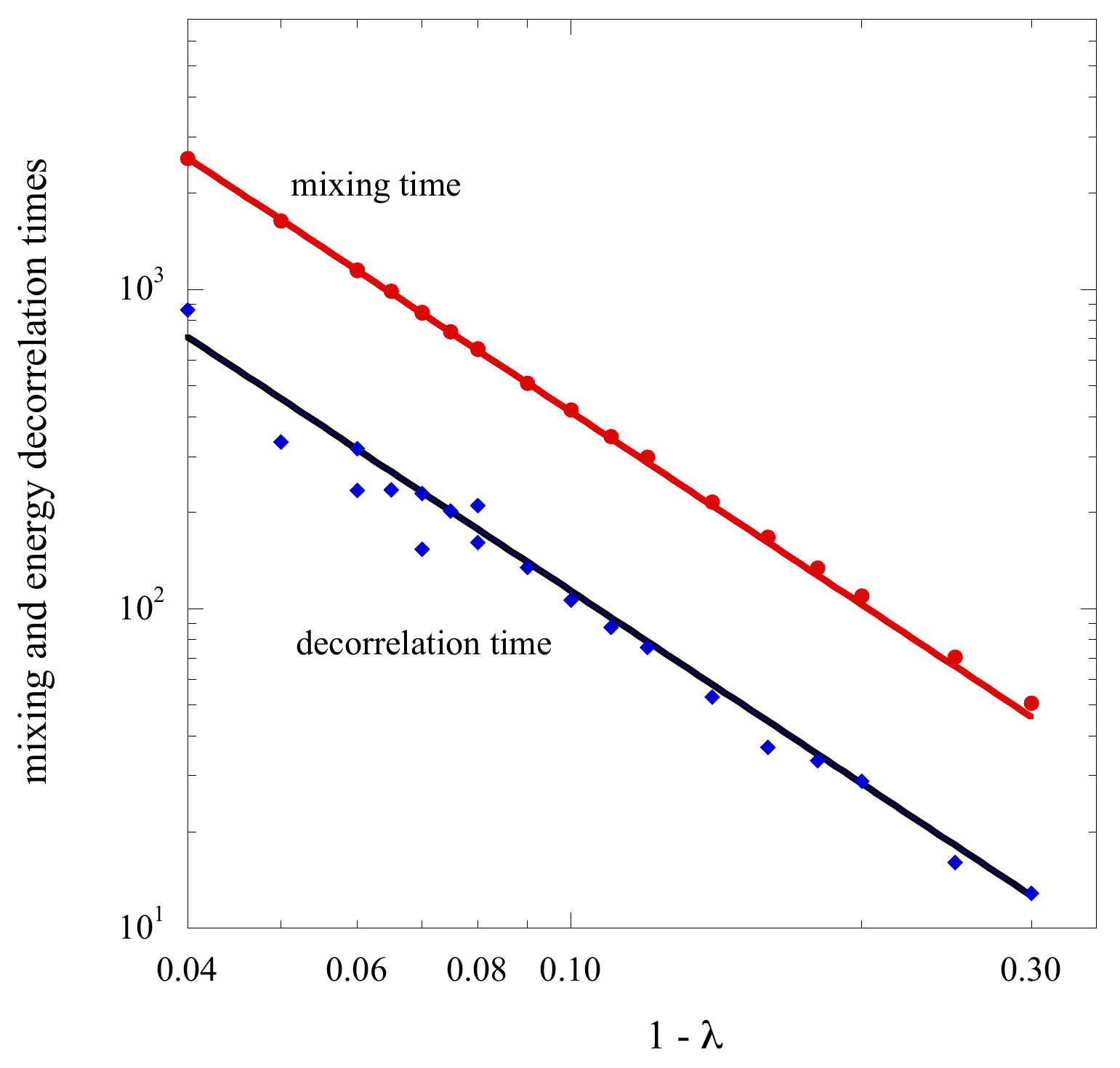}
\vspace{-0.2in}
\caption{\label{fig:lifetimeG}(a) The $\e$-dependence for constant Ginzburg parameter ($G=10^6$) of $\trich$, the lifetime of the richest agent, diverges as $\tau_{\rm ra} \sim \e^{-0.9}$ (red line). (b) The $\e$-dependence of the mixing time $\tau_m$ (top red line) and $\tau_E$ (bottom blue line) diverge as $\sim \e^{-2}$. Least squares fits give an exponent in the range $[1.95, 2.1]$.}
\end{figure}

For fixed $G$ the simulation results  for the $\e$-dependence of $\trich$ are shown in Fig.~\ref{fig:lifetimeG}(a) and are consistent with the power law $\e^{-1}$. 
To obtain accurate results for the wealth metric $\Omega(t)$, we averaged $\Omega(t)$ over ten origins and found that the linear dependence of $\Omega(0)/\Omega(t)$ holds over a wide range of $t$ and yields robust values of $\tau_m$. The exponential dependence of $C_E(t)$ holds for $t \lesssim 5 \tau_E$, yielding some uncertainty in the fitted values of $\tau_E$. The $\e$-dependence of $\tau_m$ and $\tau_E$ are shown in Fig.~\ref{fig:lifetimeG}(b). We see that both $\tau_m$ and $\tau_E$ increase rapidly as $\e \to 0$ and that their $\e$-dependence is consistent with $\e^{-2}$.

Although the mean-field theory~\cite{kleinmf} makes no direct predictions for the mixing time, the $\e$-dependence of $\tau_m$ can be understood 
by noting that the metric measures the time it takes for the average wealth of each agent
to equal the global average. Because the time it takes for the richest agent to cease being the richest and for another agent to assume that role diverges as approximately $\epsilon^{-1}$, the time for a system of $N$ agents to mix is $N\epsilon^{-1}$. Because $N \propto \e^{-1}$ for fixed $G$ [see Eq.~\eqref{eq:G}], we find that $\tau_m \sim \e^{-2}$ in agreement with the simulations.

The \mft\ of Ref.~\cite{kleinmf} predicts that the decorrelation time for fixed $G$ scales as
\be
\tau_{\rm mf} \sim \frac{1}{\mu_0 \e}. \label{eq:taumf}
\ee
The reason for the apparent discrepancy between the $\e$-dependence predicted by Eq.~\eqref{eq:taumf} and our simulation results for $\tau_m$ and $\tau_E$ is that the time unit in the simulations corresponds to $N$ exchanges, during which one agent exchanges wealth with only one agent on the average. In contrast, the applicability of mean-field theory requires that in one time unit, one agent exchanges wealth with $N$ agents on the average. Hence, the simulation and mean-field time units differ by a factor of $N$ or $1/\epsilon$ for fixed Ginzburg parameter.

The divergent behavior of $\tau_m$ and $\tau_E$ are examples of critical slowing down, which is associated with a cooperative effect and is not a property of a single agent. In contrast, $\trich$ is a property of a single agent rather than of the system as a whole and becomes independent of $\e$ if we define the time as required by the applicability of mean-field theory. 

Although the results of our simulations are consistent with the $\e$-dependence in Eq.~\eqref{eq:taumf} for constant Ginzburg parameter, Eq.~\eqref{eq:taumf} also predicts that $\tau_E$ is independent of the values of $f$ and $N$. As discussed in Ref.~\cite{kleinmf}, the derivation of Eq.~\eqref{eq:taumf} neglects the effects of both additive and multiplicative noise. The weak dependence of $\tau_m$ and $\tau_E$ on $N$ and $f$, as well as their dependence on $\mu$, is discussed in Ref.~\cite{kleinmf}.
 
\section{\label{sec:discussion}Discussion}
We have generalized the 
Yard-Sale model to incorporate economic growth and its distribution according to the wealth of the agents as determined by Eq.~\eqref{distributionofgrowth} and the parameter $\lam$. Our numerical results suggest that there are two phases. For $\lam < 1$ the system reaches a steady state with economic mobility, is effectively ergodic, and can be considered to be in thermodynamic equilibrium. In contrast, for $\lam \geq 1$ there is no economic mobility, the system does not reach a steady state, and in the limit $t \to \infty$, there is condensation of a finite fraction of the system's wealth in a vanishingly small number of agents. In addition, the system is not ergodic and shares some of the characteristics of the geometric random walk which is also not ergodic and cannot be treated by equilibrium methods~\cite{ole, ole-bill}.

It is remarkable that it is possible to define a thermodynamic energy for a system that involves wealth and has no obvious energy analogue. The interpretation of the energy and its variance is subtle and thermodynamic consistency is achieved only if the \mf\ limit is taken appropriately. 

We showed in Sec.~\ref{sec:equilibrium} that $P(E)$, the probability density of the energy of the system, is well fit by a Gaussian function for $N=5000$ and $\lam=0.8$. Simulations for $N=5000$ and values of $\lam$ much closer to one show departures from a Gaussian, even though the wealth fluctuation metric still indicates that the system is effectively ergodic. The deviation of $P(E)$ from a Gaussian for fixed $N$ (and fixed $\mu$ and $f$) is due to the fact that $G$ decreases as $\lam \to 1$ and eventually becomes too small for mean-field theory to be applicable. Simulations for fixed Ginzburg parameter begin to show deviations from a Gaussian distribution for $\lam$ much closer to one. Although the Ginzburg parameter is large and fixed, the importance of multiplicative noise increases as $\lam \to 1$, and eventually the effect of the multiplicative noise can no longer be ignored~\cite{kleinmf}.

Our numerical values of the various exponents are consistent with the \mft\ of Ref.~\cite{kleinmf}, but their estimated numerical values must be viewed with caution because they are obtained by extrapolation over a limited range of $\lam$ and for finite values of $G$ and $N$. Much larger values of $G$ would be needed to obtain more accurate numerical results for $\lam$ closer to one. 

\begin{figure}[tbp] 
\includegraphics[scale=0.6]{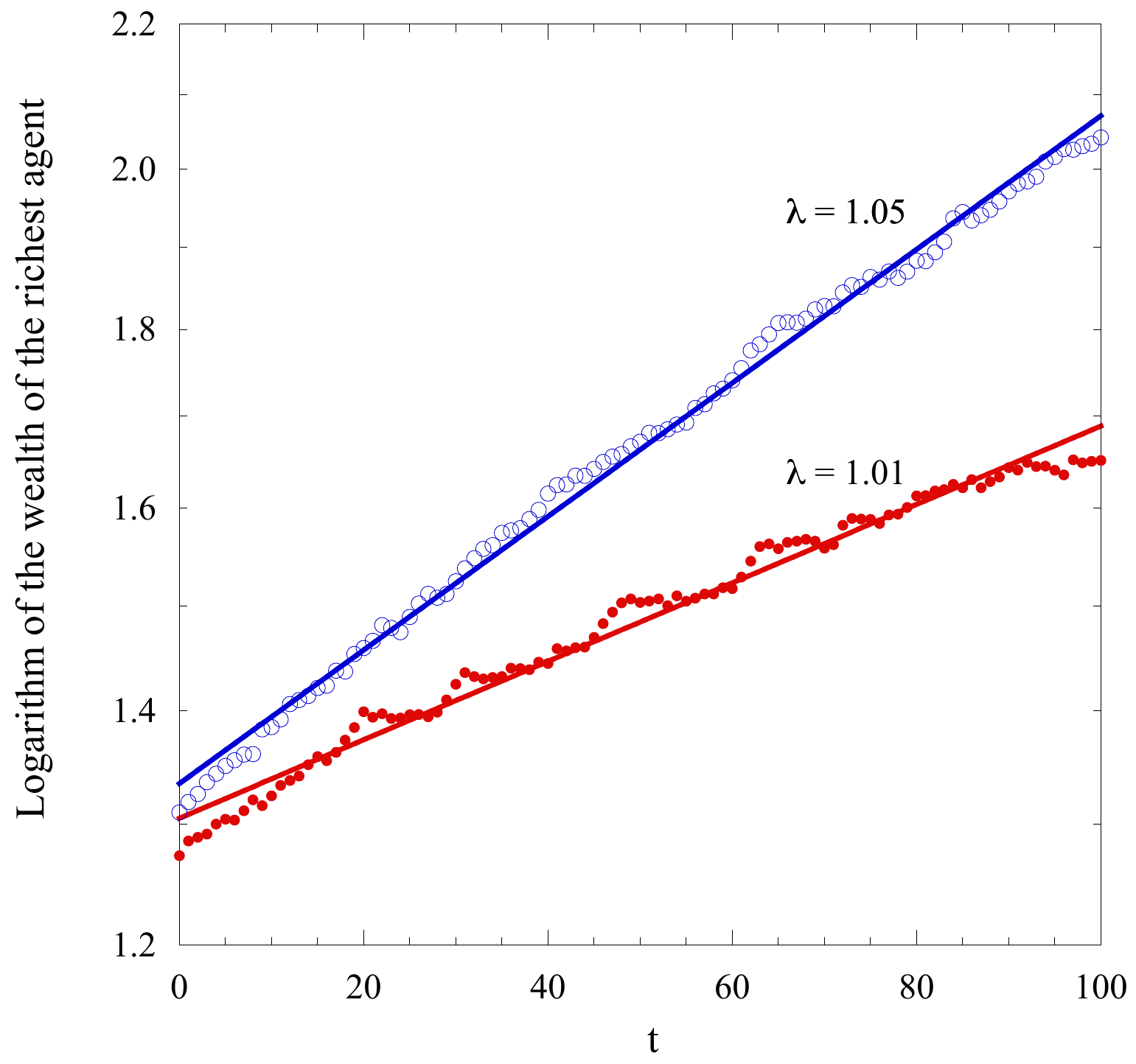}
\vspace{-0.3cm}
\caption{\label{fig:superquench}The time dependence of the wealth of the richest agent after an instantaneous change of $\lam$ from $\lam=0.8$ to $\lam=1.01$ (red) and from $\lam = 0.8$ to $\lam=1.05$ (blue).  Each change is averaged over five runs. The time $t$ is measured from the instantaneous change of $\lam$ after the system has reached equilibrium.  The wealth increases exponentially as $e^{t/\tau_q}$, with $\tau_q \approx 390$ for $\lam =1.01$ and $\tau_q \approx 230$ for $\lam = 1.05$, consistent with unstable state evolution~\cite{kleinmf} ($N=20000$, $f=0.01$, and $\mu=0.1$).}
\end{figure}

The transition at $\lam = 1$ is from a system in thermodynamic equilibrium for $\lam < 1$ to a system that undergoes wealth condensation for $\lam \geq 1$. In Ref.~\cite{kleinmf} it is shown that the evolution of the model for $\lam \geq 1$ is the same as unstable state evolution in the fully connected Ising model for model A dynamics~\cite{hh}. In Fig.~\ref{fig:superquench} we show the evolution of the wealth of 
the richest agent after the value of $\lam$ has been changed from $\lam=0.8$ to $\lam=1.01$ and to $\lam=1.05$. We see that the wealth of the richest agent initially increases exponentially. We also find that the duration of exponential growth decreases for larger values of $\lam$ after the change (not shown).

Our results have possible important economic implications. 
As $\lam$ is increased, the benefits of growth are weighted more toward the wealthy, and wealth inequality increases. Nevertheless, as long as $\lam < 1$, the wealth of agents of all ranks grows at the same rate once a steady state is reached, and all agents benefit from economic growth. However, if the benefits of growth are skewed too much toward the wealthy ($\lam \geq 1$), poor and middle rank agents no longer benefit from economic growth, and wealth condensation occurs. For $\lam=1$ the model reduces to the geometric random walk with resulting wealth condensation~\cite{ole}, as shown in the following paper~\cite{kleinmf}.

There is some question whether economic systems can be treated as being in equilibrium or even exhibit effective ergodicity~\cite{ole, ole-bill, rmp}. Our results suggest that ergodicity and equilibrium may depend on various system parameters.
Because parameters such as $\lam$ and $\mu$ are not temporal constants in real economies, our results also suggest that the applicability of equilibrium methods may be situational and vary with time.

\begin{acknowledgments}
We thank Timothy Khouw, Ole Peters, John Ogren, Alan Gabel, Bill Gibson, Karen Smith, and Louis Colonna-Romano for useful conversations.
\end{acknowledgments}

\appendix*

\section{Comparison to some economic data}

Modeling the economy of a country as large and diverse as the United States by compressing economic growth and transactions into three parameters, $\lambda$, $f$, and $\mu$ is a gross simplification. The assumption that these parameters are independent of time also is unrealistic. In the following, we analyze the growth data~\cite{smith1} and wealth distribution data~\cite{smith2} compiled by Karen Smith and published by the Urban Institute. Our analysis suggests that the assumptions that the distribution of growth can be modeled as in Eq.~\eqref{distributionofgrowth} and that the parameters are independent of time is a
reasonable {\it zeroth order approximation} to the distribution of wealth in the real economy. We will discuss the exchange term and its relevance to the real economy in the following paper~\cite{kleinmf}.

From the growth rate of the gross domestic product shown in constant dollars in Ref.~\cite{smith1}, we note that the temporal fluctuations of the (inflation adjusted) growth rate of the gross domestic product exhibits large swings that appear to be damped as a function of time. The decline in the growth caused by the great recession starting in 2008 is an example of a large fluctuation, but the growth rate has remained close to the mean rate of roughly 3\% over the last 30 years, thus implying that $\mu \approx 0.03$.

By using the wealth distribution chart in Ref.~\cite{smith2}, we can calculate the change of the wealth of people in various percentiles. 
From the relation [see Eq.~\eqref{distributionofgrowth}], we can estimate $\lambda$ as
\be
\lam = \log \Bigg( \frac{W_{r}(t_{2}) - W_{r}(t_{1})}{W_r(t_1)} \Bigg),
\ee
where $W_{r}(t)$ is the wealth of people of economic rank (percentile) $r$ at time $t$.

We estimated $\lambda$ for the 50th, 90th and 95th wealth percentiles in the intervals 1983--1989, 1995 --1998 and 2013--2016 (see Table~\ref{tab:lambda}). Although the values of $\lambda$ are not constant for different time intervals and percentiles, they vary by only a few 
percentage points as a function of percentile. They vary more as a function of time, with the 50\% percentile having the greatest variation. The change of $\lam$ appears to decrease for later times, consistent with the damping of the variation of the growth.
The variation of $\lambda$ is more pronounced for even lower rankings. However, because the wealth of the lower rankings is considerably smaller, the variation of the value of $\lambda$ has less effect on the wealth of the poor.

\begin{table}[t]
\begin{tabular}{|c|c|c|c|}
\hline
percentile & 1983--1989 & 1995--1998 &2013--2016\\
\hline
95\% & 0.85 & 0.90 & 0.90 \\
\hline
90\% & 0.84 & 0.89 & 0.90\\
\hline
50\% & 0.75 & 0.90 & 0.84 \\
\hline
\end{tabular}
\caption{\label{tab:lambda}The calculated values of $\lambda$ for the percentiles and time intervals indicated using economic data from Refs.~\cite{smith1} and \cite{smith2}.}
\end{table}

We conclude from the growth and wealth distribution data~\cite{smith1,smith2} that the distribution of economic growth assumed in the GED model is a reasonable zeroth order approximation, particularly for the upper half of the wealth ranks of the United States over the past 30 years. Of course, there is much that it is not included in the model, such as the effects of wars, famines, storms, and recessions, which are not obtainable from the simple GED model. However, it appears from the data that the model is a reasonable approximation over time scales of the order of decades and yields insights into the importance of how economic growth is distributed.

\end{document}